\documentclass[aps,prx,twocolumn,superscriptaddress,longbibliography,nofootinbib,amsmath,amssymb]{revtex4-2} 
\usepackage[T1]{fontenc}
\usepackage{subcaption} 
\usepackage{graphicx} \graphicspath{ {./figs/} }
\usepackage{physics}
\usepackage{dsfont}
\usepackage{bm}
\usepackage[dvipsnames]{xcolor}
\usepackage{hyperref}
\usepackage{tikz}
\usepackage{mathrsfs}
\usetikzlibrary{calc}
\usetikzlibrary{knots}
\usepackage{todonotes}
\usepackage{enumerate}
\usepackage{ragged2e}
\usepackage{amsthm}
\usepackage[normalem]{ulem}
\usepackage{cancel}

\makeatletter 
\renewcommand\onecolumngrid{
\do@columngrid{one}{\@ne}%
\def\set@footnotewidth{\onecolumngrid}
\def\footnoterule{\kern-6pt\hrule width 1.5in\kern6pt}%
}

\renewcommand\twocolumngrid{
        \def\footnoterule{
        \dimen@\skip\footins\divide\dimen@\thr@@
        \kern-\dimen@\hrule width.5in\kern\dimen@}
        \do@columngrid{mlt}{\tw@}
}%

\def\amsbb{\use@mathgroup \M@U \symAMSb}
\makeatother
\DeclareMathOperator{\parity}{parity}

\newtheorem{lemma}{Lemma}

\newtheorem{definition}{Definition}

\newcommand{\ie}{\emph{i.e.\@} }
\newcommand{\eg}{\emph{e.g.\@} }
\newcommand{\rme}{\mathrm{e}}

\def\cell{0.375} 
\newcommand{\SU}{\mathrm{SU}}

\begin{document}
\title{Typical entanglement in anyon chains: Page curves beyond Lie group symmetries}
\author{Yale Yauk}
\email{yale.yauk@mpq.mpg.de}
\affiliation{Max-Planck-Institut für Quantenoptik, 85748 Garching, Germany}
\affiliation{Munich Center for Quantum Science and Technology (MCQST), 80799 München, Germany}

\author{Lucas Hackl}
\affiliation{School of Mathematics and Statistics, The University of Melbourne, Parkville, Victoria 3010, Australia}
\affiliation{School of Physics, The University of Melbourne, Parkville, Victoria 3010, Australia}

\author{Alexander Hahn}
\email{alexander.hahn@mpq.mpg.de}
\affiliation{Max-Planck-Institut für Quantenoptik, 85748 Garching, Germany}
\affiliation{TUM School of Natural Sciences, Technical University of Munich, 85748 Garching, Germany}
\affiliation{Munich Center for Quantum Science and Technology (MCQST), 80799 München, Germany}

\begin{abstract}
We study bipartite entanglement statistics in one-dimensional anyon chains, whose Hilbert spaces are constrained by fusion rules of unitary pre-modular categories. Our setup generalizes previous frameworks on symmetry-resolved entanglement entropy for non-abelian Lie group symmetries to the setting of quantum groups. We derive analytical expressions for the average anyonic entanglement entropy and its variance. Surprisingly, despite the constrained Hilbert space structure, the large $L$ expansion has no universal $O(\sqrt{L})$ or $O(1)$ symmetry-type corrections except for a subleading topological correction term that produces a Page curve asymmetry. We further show that the variance decays exponentially with system size, establishing the typicality. Numerical simulations of the integrable and quantum-chaotic golden chain Hamiltonian show that chaotic mid-spectrum eigenstates match the Haar-random predictions, supporting the use of eigenstate entanglement as a diagnostic of quantum chaos. Our results establish the anyonic Page curve as an appropriate chaotic benchmark in topological many-body systems and connect anyonic entanglement to Page-type universality in quantum many-body physics.
\end{abstract}

\maketitle

\clearpage

\section{Introduction}
Entanglement is a defining feature of quantum many-body systems~\cite{TEE1,TEE2, Vidal2003, Calabrese2004}, a central concept in quantum information theory~\cite{Jozsa2003, Gottesman1997, Bennett1993, Bennett1992, Zukowski1993, Ekert1991}, and plays an important role in high-energy physics~\cite{PageBH} and holography~\cite{RT}\@.
Given this broad relevance, it is natural to ask what amount and structure of entanglement one should expect generically.
This motivates the study of the average entanglement entropy (EE) of bipartite Haar-random pure states.
Techniques developed in random matrix theory have enabled analytical calculations of the average EE of Haar-random states, such as the seminal result obtained by Page \cite{page_curve}\@. It says that the average EE is nearly maximal, deviating by a term surviving only at half-bipartition in the thermodynamic limit.

The average EE also serves as an important baseline for the entanglement of particular physically relevant states: 
It is well-established that ground states and low-lying energy eigenstates of local Hamiltonians satisfy an area law for entanglement~\cite{arealaw1, arealaw2}\@. In contrast, the entanglement of highly excited energy eigenstates typically obey a volume law. In particular, the bipartite EE behaves very differently in quantum-chaotic~\cite{typicalEEreview,eigenstatechaos,eigenstatechaos2,averageEEchaos1,averageEEchaos2,averageEEchaos3,averageEEchaos4,chaosKLdivergenceEE,averageRenyichaos}, integrable-interacting~\cite{XXXEE,XXZEE,LMGEE}, and Gaussian quadratic~\cite{gaussianfermion1,gaussianfermion2,gaussianfermion3,gaussianfermion4,gaussianfermion5,gaussianfermion6,gaussianfermion7,gaussianboson,gaussianboson2} models. This makes entanglement a powerful diagnostic for distinguishing universality classes of many-body systems. In this context, Haar-random entanglement provides an especially useful benchmark: it has been observed that the Haar-random average EE agrees with the EE of eigenstates of local quantum-chaotic Hamiltonians up to a constant $O(1)$ term~\cite{eigenstatechaos,averageEEchaos4,averageEEchaos3,chaosKLdivergenceEE}\@. This suggests that, to leading order, eigenstates for these models resemble typical states. It is thus conjectured that the average EE of highly-excited eigenstates is a universal (model-independent) diagnostic of quantum chaos, though interesting counterexamples can be found in systems with strong dynamical constraints~\cite{area_law_EE_chaotic}\@.

In realistic many-body systems, however, global symmetries impose additional constraints. Recently, there has been considerable interest in understanding the role of Lie group symmetries on the EE, both from the Hamiltonian and the Haar-random perspectives. For abelian $\mathrm{U}(1)$ symmetries (\eg particle number or spin magnetization conservation), the Haar-random average EE exhibits subleading corrections to the Page curve at order $O(\sqrt{L})$ (where $L$ is the size of the system) and $O(1)$~\cite{averageEEchaos1,typicalEEreview,ETHU(1),EEU(1)1,EEU(1)2}. The case for non-abelian $\SU(2)$ symmetries in spin chains has also been studied~\cite{EESU(2)2,EESU(2)1,SREESU(2),EESU(2)1.1}, where similar conclusions were drawn. More generally, for a non-abelian compact Lie group $G$, one can define a symmetry-resolved EE, which quantifies the entanglement of $G$-invariant degrees of freedom in the system. In this work, we show that anyonic entanglement provides the natural quantum group analogue of this construction, thus extending the study of typical entanglement beyond Lie group symmetries.

The symmetry-constrained setting highlights a broader principle---that typical entanglement is sensitive to the structure of the Hilbert space from which states are drawn. A paradigmatic example of a system with rich intrinsic constraints arises in topologically ordered phases, whose emergent quasiparticles---anyons---obey exchange statistics beyond bosons and fermions and are described by unitary representations of the braid-group ~\cite{anyon_statistics_1,anyon_statistics_2}\@. Such systems are characterized by long-range entanglement and topological robustness, making them a promising platform for quantum computing~\cite{topological_quantum_computing_review,TEE1,TEE2}\@. The Hilbert space of non-abelian anyons is not a simple tensor product but is restricted by fusion rules and carries built-in superselection sectors~\cite{superselection} of topological charge~\cite{Kitaev_anyon, Bonderson_thesis, Kong2022}. This raises a basic foundational question:
\emph{What is the typical entanglement structure of topologically constrained Hilbert spaces, and do Page-like universality statements survive in non-abelian topological phases?}

In this work, we address this question by deriving analytical formulas for the Haar-average bipartite EE and its variance. In contrast to Lie group symmetries, we recover the Page curve without any $O(\sqrt{L})$ or $O(1)$ corrections. Instead, we find a topological-sector correction that leads to an asymmetry of the Page curve if the total anyonic charge is non-abelian. We further show that the variance of the entropy decays exponentially with system size, establishing the typicality of the result. Finally, we verify these predictions numerically in a quantum-chaotic anyonic Hamiltonian, where mid-spectrum eigenstates match the Haar-random entanglement structure. This is consistent with earlier indications of eigenstate thermalization in anyonic systems~\cite{ETHanyons}\@. In this way, our results connect the anyonic entanglement program with the broader theme of Page-type typicality and universality in many-body quantum systems.

The paper is organized as follows. We start by reviewing the basic concepts of anyonic quantum information in Section~\ref{sec:anyonic_quantum_info}\@. 
We show that our results are the quantum group generalization of the non-abelian symmetry-resolved entanglement in Section~\ref{sec:symmetry-resolved}\@. Afterwards, we present our main analytical results of the average anyonic entanglement entropy and its variance in Section~\ref{sec:average_and_variance}\@. 
Section~\ref{sec:eigenstate_AEE} is dedicated to studying eigenstate entanglement of integrable and quantum-chaotic anyonic Hamiltonians and its relation to the Haar-random prediction. Finally, we conclude in Section~\ref{sec:conclusion}\@. For all mentions of category theory and superselection, the relevant definitions can be found in Appendix~\ref{sec:anyon_basics}\@. Appendix~\ref{app:variance} outlines the derivation of the variance. Furthermore, we compute the non-anyonic $U_q(\mathfrak{sl}_2)$-symmetry-resolved entanglement entropy in Appendix~\ref{app:vN_entropy} to compare to both our anyonic and the Lie group case.

\section{Anyonic quantum information}\label{sec:anyonic_quantum_info}
In this section, we review standard information-theoretic concepts, applied to anyonic systems. We then introduce the anyonic entanglement entropy (AEE), a measure of bipartite entanglement for anyonic pure states. We assume familiarity with basic anyonic Hilbert space and the diagrammatic fusion-tree notation, often referring to a minimal review provided in Appendix~\ref{sec:anyon_basics}.

\subsection{Setup and conventions}\label{sec:setup}
Anyonic charges are superselection sector labels from a finite set $\mathcal{S}$, which are the simple objects in a unitary modular tensor category (UMTC) or, more generally, in a unitary premodular category (UPMC). See Appendix~\ref{app:category_theory} for a brief discussion on these categories and Appendix~\ref{app:superselection} for superselection. These conserved charges satisfy the fusion algebra $a\otimes b=\bigoplus_{c\in\mathcal{S}} N_{ab}^c c$, where $N_{ab}^c\in\mathbb{N}$ are the fusion multiplicities. For simplicity, we will only consider the multiplicity-free case, \ie $N_{ab}^c\in\{0,1\}$, for all $a,b,c\in\mathcal{S}$. The fusion of two anyons defines a fusion space $V_{ab}^c$ of dimension $N_{ab}^c$. 

We consider $L$ anyons $a_1,\dots,a_L$ on a disk\footnote{This fixes the topology to a simply connected surface, where the absence of non-contractible loops eliminates additional topological constraints. Consequently, the Hilbert space is determined entirely by the anyonic fusion degrees of freedom. For a discussion on higher-genus surfaces, see Appendix~\ref{app:PBC}\@.} which fuse to a total charge $J$, forming an anyon chain. Throughout, we restrict to the setting where all anyons on the chain are identical, \ie $a_1=\cdots=a_L=\mathfrak{j}$. Then, the total Hilbert space decomposes as
\begin{equation}
    \mathcal{H}=\bigoplus_{J\in\mathcal{S}}\mathcal{H}^J\,, \quad \mathcal{H}^J=V^J_{\mathfrak{j}^{\times L}}\,,
\end{equation}
where $V^J_{\mathfrak{j}^{\times L}}$ is the fusion space of $L$ anyons of charge $\mathfrak{j}$ (which could be trivial if the fusion is disallowed). We denote the dimension of the fusion space by $D_J(L)=\dim \mathcal{H}^J$, leaving a lower-case $d_J$ for the quantum dimension of charge $J$. A canonical basis of $\mathcal{H}^J$ is given by vectors $\ket{\vb{x};J}$, where the vector $\vb{x}$ labels the intermediate fusion labels. Due to superselection, the algebra of observables $\mathcal{M}$ consists only of those which preserve the total charge $J$, $\mathcal{M}=\bigoplus_J \mathcal{B}(\mathcal{H}^J)$. 

Adopting the notation of \cite{Bonderson_entanglement}, we define an anyonic density operator $\tilde{\rho}$ representing a state to be a semi-positive definite operator normalized with respect to the quantum trace, $\tilde{\tr}\tilde{\rho}=1$. We denote the space of states by $\mathscr{S}(\mathcal{M})$ and will use this notation interchangeably for both the positive linear functionals on $\mathcal{M}$ and their representations as density operators. These conventions ensure that the state and operations on it are compatible with isotopy-invariance required by the theory, as discussed in Appendix~\ref{app:anyon_diagrams}.

An anyonic charge $a$ is called abelian if, for every $b\in\mathcal{S}$, the fusion outcome is unique in the sense that $\sum_c N_{ab}^c = 1$. Otherwise, it is called non-abelian. There is an ongoing discussion in the literature~\cite{anyon_asymmetry,superactivation_anyon} about whether states $\tilde{\rho}=1/d_J\ketbra{J}{J}$ with non-abelian total charge $J$ are pure, where $\ket{J}=\sum_{\vb{x}} c_{\vb{x}}\ket{\vb{x};J}$. This issue is pertinent to us because we study the entanglement properties of such states, and pure versus mixed states can exhibit fundamentally different entanglement behavior. We offer a different perspective on this debate based on operator algebras. This leads us to distinguish three different classes of states~\cite{Kato_entanglement}: i) pure, with abelian total charge $J$; ii) pure, with non-abelian total charge $J$; iii) mixed. See Appendix~\ref{app:pure_states} for details. This motivates the following definition.
\begin{definition}
    A state $\tilde{\rho}\in\mathscr{S}(\mathcal{M})$ is said to be pure if it is extremal, \ie if $\tilde{\rho}=\lambda\tilde{\rho}_1+(1-\lambda)\tilde{\rho}_2$ for $0<\lambda<1$ and $\tilde{\rho}_1,\tilde{\rho}_2\in\mathscr{S}(\mathcal{M})$ implies that $\tilde{\rho}_1=\tilde{\rho}_2$. Equivalently, $\tilde{\rho}$ is supported on a definite charge sector $J$ and is a rank-1 projector, \ie there exists $\ket{J}\in \mathcal{H}^J$ such that $\tilde{\rho}=1/d_J\ketbra{J}{J}$.
\end{definition}
We note that this definition of purity differs from that of Ref.~\cite{Bonderson_entanglement}, because states with non-abelian total charge can be pure as proved in Lemma~\ref{lem:pure_states}\@.

\subsection{Examples}
\label{sec:anyon_examples}
We now introduce some explicit examples of anyon theories that arise in physical systems.

\subsubsection{Abelian anyons}
An anyon theory is abelian if all its charges are abelian. One prominent example is the quantum double $D(\mathbb{Z}_N)$ of $\mathbb{Z}_N$, which is a UMTC describing several physical settings, such as orbifold conformal field theories~\cite{Dijkgraaf1989, Dijkgraaf1991}, topological $\mathbb{Z}_N$ gauge theories~\cite{Bais1992} and the $\mathbb{Z}_N$ toric code~\cite{toric_code}. Other examples are the UPMC $\mathrm{Rep}(\mathbb{Z}_2)$, which describes global bosonic $\mathbb{Z}_2$ symmetries~\cite{Thorngren2024}, or the premodular category of super vector spaces $\mathrm{sVec}$, which is related to fermionic topological order~\cite{Bruillard2017, Cho2023}\@. 

For abelian models, the set $\mathcal{S}$ of anyonic charges forms an abelian group under the fusion. That is, $a\otimes b = a\cdot b$ for a unique simple object $a\cdot b\in\mathcal{S}$, and thus $N_{ab}^c = \delta_{c, a\cdot b}$. All fusion spaces are trivial, in the sense that if they exist, they are one-dimensional and the fusion path is unique. As a consequence, it can be shown that $d_a=1$ for all $a\in\mathcal{S}$. 

\subsubsection{$\SU(2)_k$ anyons}
The $\SU(2)_k$ theories provide a canonical example of non-abelian topological phases. They are algebraically described by the UMTC $\mathrm{Rep}(U_q(\mathfrak{sl}_2))$ with deformation parameter $q=\rme^{\frac{2\pi i}{k+2}}$. This same UMTC defines a $(2+1)D$ topological quantum field theory (TQFT), where $\SU(2)_k$ anyons arise as excitations of an $\SU(2)$ Chern--Simons theory at level $k$ \cite{Witten}. This theory has been used to successfully describe certain quasiparticle excitations in fractional quantum Hall states \cite{FQH}, which exhibit exotic non-abelian braiding statistics.

Throughout this work, $\SU(2)_k$ anyons will serve as our guiding example. The simple objects are labeled by finitely many spins $\mathcal{S}=\{0,1/2,1,\dots,k/2\}$ and given by fusion rules $j_1\otimes j_2=\bigoplus_{j=\abs{j_1-j_2}}^{\min(j_1+j_2,k-j_1-j_2)}j$. All objects in $\SU(2)_k$ are self-dual, $j=\overline{j}$. For obvious reasons, the $\SU(2)_k$ anyon theories are often informally thought of as $\SU(2)$ theories with a spin cut-off at $k/2$. We will interchangeably call the charge $J\in\mathcal{S}$ of a collection of anyons the total spin, as is more common when viewing them as deformations of $\SU(2)$ spins.

\subsubsection{Fibonacci anyons}
The Fibonacci anyon model, with associated UMTC denoted by $\mathrm{Fib}$, is one of the simplest and most widely studied non-abelian anyon models. It contains only two anyon types $\mathcal{S}=\{0,\tau\}$, with the only non-trivial fusion rule being $\tau\otimes\tau=0\oplus\tau$. Actually, the Fibonacci anyons form the integer charge (half-integer) charge sector of $\SU(2)_3$, under the identification $0\simeq 0$\,($3/2$) and $\tau\simeq 1$\,($1/2$), so one often informally writes $\SU(2)_3=\mathrm{Fib}\times \mathrm{Fib}$ (or formally: $\SU(2)_3=\mathrm{Fib}\rtimes\mathbb{Z}_2$).

A major reason for the continued interest in Fibonacci anyons is the fact that the model supports universal quantum computation \cite{topological_quantum_computing_review}. More precisely, just from a series of elementary braids, one can approximate any unitary gate to arbitrary precision. Furthermore, anyonic quantum computing enjoys the property of being robust to noise, as perturbations that do not affect the topology of the braiding process do not affect the logical computation.

\subsection{Anyonic entanglement entropy}\label{sec:AEE}
Although anyonic systems lack a notion of a local Hilbert space, \textit{i.e.}\@, any subsystem of a single anyon is trivial ($1$-dimensional), a bipartition of $L$ anyons into two sets $A=\{a_1,\dots,a_{L_A}\}$ and $B=\{a_{L_A+1},\dots,a_L\}$ containing $L_A>1$ and $L_B=L-L_A$ anyons is well-defined. Specifically, the Hilbert space can be decomposed as to respect the bipartition:
\begin{equation}
\label{eq:bipartite_fusion_space}
    \mathcal{H}^J_{AB}=\bigoplus_{\alpha,\beta} \left(\mathcal{H}^\alpha_A\otimes \mathcal{H}^\beta_B\otimes V_{\alpha\beta}^J\right)\,, 
\end{equation}
where $\mathcal{H}^\alpha_A=V^\alpha_{\mathfrak{j}^{\times L_A}}$ and $\mathcal{H}^\beta_B=V^\beta_{\mathfrak{j}^{\times L_B}}$. We denote vectors in $\mathcal{H}_{AB}^J$ with $\ket{\vb{x},\alpha;\vb{y},\beta;J}$, where $\vb{x}$ are the intermediate fusion labels on side $A$, $\alpha$ is the total charge on side $A$, $\vb{y}$ are the intermediate fusion labels on side $B$ and $\beta$ is the total charge on side $B$. This allows us to define separable anyonic states, following Ref.~\cite{Kato_entanglement}\@.
\begin{definition}
    An anyonic pure state $\ket{\vb{x},\alpha,\vb{y},\beta;J}\in\mathcal{H}_{AB}^J$ is called separable, if (i) there exist $\ket{\vb{x};\alpha}\in\mathcal{H}_A^\alpha$ and $\ket{\vb{y};\beta}\in\mathcal{H}_B^\beta$, such that $\ket{\vb{x},\alpha,\vb{y},\beta;J}$ can be written as $\ket{\vb{x},\alpha,\vb{y},\beta;J}=\ket{\vb{x};\alpha}\otimes\ket{\vb{y};\beta}$, and (ii) at least one of the total charges $\alpha,\beta$ is abelian, so that $V_{\alpha\beta}^J$ is one-dimensional. A mixed anyonic state $\tilde{\rho}$ on $\mathcal{H}_{AB}^J$ is called separable if it can be written as a convex combination of separable pure anyonic states. States, which are not separable, are called entangled. 
\end{definition}

A faithful entanglement measure has only been defined for those pure states with an abelian total charge \cite{Hikami_entanglement,Pfeifer_entanglement,Bonderson_entanglement}. This anyonic entanglement entropy (AEE) is the diagrammatic generalization of the von Neumann entropy, given by 
\begin{equation}
\label{eq:AEE_definition}
    \tilde{S}_A=-\tilde{\tr}(\tilde{\rho}_A\log \tilde{\rho}_A)\,, \quad \tilde{\rho}_A=\tilde{\tr}_B\tilde{\rho}\,. 
\end{equation}
In practice, the AEE is calculated using the replica trick $\tilde{S}_A=-\partial_r \tilde{\tr}\tilde{\rho}_A^r\big\rvert_{r=1}$. It measures contributions to entanglement additionally arising from the topological degrees of freedom in the system.

An equivalent formulation of the AEE (which the authors called the asymptotic entanglement entropy) was provided in Ref.~\cite{Kato_entanglement}, motivated by the operational meaning of the von Neumann entropy in the limit of many copies. Indeed, the quantum trace is the operation compatible with tensor products of copies, so the asymptotic entanglement entropy has the same form as~\eqref{eq:AEE_definition}.

Strictly speaking, the AEE is not an exact measure of entanglement for states with non-abelian charge, because, as mentioned in the previous paragraphs, these states contain information and exhibit mixed-state entanglement properties. For example, it can be shown that for states with non-abelian total charge $J$, the entanglement entropy is asymmetric under subsystem exchange, \ie $\tilde{S}_A\neq \tilde{S}_B$. However, the AEE can still be used as a quantifier of quantum correlations. It can also be used to form the mutual information $\tilde{S}(A:B)=\tilde{S}_A+\tilde{S}_B-\tilde{S}_{AB}$, which is an exact measure for quantum and classical correlations.

Other diagnostics of anyonic entanglement have been broadened beyond the AEE include anyonic generalizations of partial transposes and logarithmic negativity, as well as studies of correlations and resource-theoretic structures constrained by topological superselection~\cite{anyonic_partial_transpose,Kirchner_entanglement,topological_correlation,anyon_asymmetry,anyonic_relative_entropy}. We do not study these measures further as they are not as amenable to analytic computation.

\subsection{Relation to symmetry-resolved entanglement}\label{sec:symmetry-resolved}
We elucidate a connection between anyonic entanglement and the non-abelian symmetry-resolved entanglement studied in Ref.~\cite{SREESU(2)}. We review the notion of symmetry-resolution of a compact Lie group $G$, using $G=\SU(2)$ as the paradigmatic example. We argue that the anyonic entanglement entropy computed in this manuscript can be understood as a $q$-deformed symmetry-resolved entanglement entropy, $q$ being the deformation parameter of the associated quantum group $U_q(\mathfrak{sl}_2)$.

Consider a chain of $L$ spin-$1/2$ particles which has the Hilbert space $\mathcal{H}=(\mathbb{C}^{2})^{\otimes L}$ built out of tensor products of spin-$1/2$ irreducible representations of $\SU(2)$. The Schur--Weyl duality guarantees a decomposition
\begin{equation}
\label{eq:Schur-Weyl}
    \mathcal{H}=\bigoplus_J \left(\mathcal{H}^{J}_{\mathrm{sym}}\otimes \mathcal{H}^{J}_G\right)
\end{equation}
into a direct sum over the total spin $J$, and further into a tensor product between the $(2J+1)$-dimensional spin-$J$ irreducible representation $\mathcal{H}^J_{\mathrm{sym}}$ and a $G$-invariant intertwiner or multiplicity space $\mathcal{H}^J_G$, which is an irreducible representation of the symmetric group $S_L$. An equivalent perspective is that this decomposition is adapted to the subalgebra $\mathcal{M}_G$ generated by the generators of $\SU(2)$, and its commutant $\mathcal{M}_\mathrm{sym}=(\mathcal{M}_G)'$, the algebra of $G$-invariant observables.

The symmetry resolution is enforced by restricting to states which are i) supported in one $J$ sector, ii) product states with respect to the tensor product \eqref{eq:Schur-Weyl}. The $\mathcal{H}^J_{\mathrm{sym}}$ factor is treated as an ancilla, and the focus is now on states in $\mathcal{H}^J_G$, where $\SU(2)$ acts trivially. Given a spatial bipartition into subsystems $A$ and $B$, the multiplicity space admits a decomposition $\mathcal{H}^J_G=\bigoplus_\alpha \mathcal{H}^{\alpha}_{GA}\otimes \mathcal{H}^{J,\alpha}_{GB}$, where $\alpha$ labels the total spin of subsystem $A$, and $\mathcal{H}^\alpha_{GA}$ and $\mathcal{H}^{J,\alpha}_{GB}$ are multiplicity spaces compatible with total spin $J$. This exactly parallels the fusion space decomposition Eq.~\eqref{eq:bipartite_fusion_space}\@, but for $\SU(2)$. The symmetry-resolved entanglement is defined as the entanglement between these two bipartitions.

The situation is far more complicated in the anyonic setting because at root of unity $q=\rme^{\frac{2\pi i}{k+2}}$, the category of finite-dimensional $U_q(\mathfrak{sl}_2)$-modules is not semisimple. For quantum spin chains such as the XXZ chain, $U_q(\mathfrak{sl}_2)$ acts on $(\mathbb{C}^2)^{\otimes L}$ by comultiplication, and the commutant algebra is a representation of the Temperley--Lieb algebra \cite{quantum_Schur_Weyl}. At $q$ root of unity, neither the $U_q(\mathfrak{sl}_2)$ nor the Temperley--Lieb structures are semisimple; the semisimple anyon theory is recovered by a process of semisimplification. Concretely,  one restricts attention to the so-called tilting modules and quotients out by the negligible modules, yielding the fusion category $\mathrm{Rep}(\SU(2)_k)$. In this semisimple category, the tensor powers of the fundamental object $V_{\frac{1}{2}}$ decomposes as 
\begin{equation}
    \mathcal{H}=V_{\frac{1}{2}}^{\otimes L}=\bigoplus_J V^J\otimes \mathrm{Hom}(V^J,V_{\frac{1}{2}}^{\otimes L})\,,
\end{equation}
where now $V^J$ are simple objects of $\mathrm{Rep}(\SU(2)_k)$ and $\mathcal{H}^J=\mathrm{Hom}(V^J,(\mathbb{C}^2)^{\otimes L})$ is precisely the anyonic fusion space of total charge $J$.

We end with a brief summary. The full mathematically rigorous treatment of the representation theory of $U_q(\mathfrak{sl}_2)$ is far beyond the scope of this work. The essential point is that anyonic Hilbert space can be identified with the quantum group symmetry-resolved multiplicity space, whose dimension is just the multiplicity of the charge label in the semisimple decomposition---or equivalently, the number of fusion paths. Therefore, the results in this work can be understood as extending previous symmetry-resolved entanglement results for compact Lie groups to the setting of quantum groups and beyond.\footnote{We note that the AEE considered here is more general, as it can also be applied to arbitrary UPMCs.} Up to the choice of entanglement measure, we compute the $q$-deformed symmetry-resolved entanglement entropy of bipartite pure states. Indeed, in Appendix~\ref{app:vN_entropy}, we calculate the average von Neumann entropy of random pure states in fusion space---this is the direct $q$-deformed analogue of the non-abelian symmetry-resolved entanglement entropy studied in Ref.~\cite{SREESU(2)}, providing a natural point of comparison.

\section{Anyonic entanglement: average and variance}
\label{sec:average_and_variance}
In this section, we analytically calculate the Haar-random average $\langle \tilde{S}_A\rangle$ over states with fixed topological charge $J$ [Eq.~\eqref{eq:SU(2)_k_AEE}] and the leading-order variance [Eq.~\eqref{eq:SU(2)_k_variance}]. Independent of the anyon model, this variance vanishes in the thermodynamic limit, so the formulas describe the typical AEE of anyonic states with fixed $J$. We also apply our results to the anyon examples listed in Section~\ref{sec:anyon_examples} and compute the large $L$ expansion.

\subsection{Average anyonic entanglement entropy}
\label{sec:average_entropy}
We begin by calculating the AEE of a general state, before averaging it over all states. Let $\tilde{\rho}=1/d_J\ketbra{\Psi}{\Psi}$ be a pure state on $\mathcal{H}^J$. Omitting the identical external anyon labels $\mathfrak{j}$, a general $\tilde{\rho}$ reads
\begin{equation}
    \label{eq:bipartition}\tilde{\rho}=\sum_{\substack{\vb{x},\alpha,\vb{y} \\ \vb{x}',\alpha',\vb{y}'}}\frac{\Psi_{\vb{x},\alpha,\vb{y}}\Psi^*_{\vb{x}',\alpha',\vb{y}'}}{\sqrt{d_J d_{\mathfrak{j}}^L}} \hspace{-0.5cm}\begin{tikzpicture}[baseline={([yshift=-0.5ex]current bounding box.center)},x=\cell cm,y=\cell cm]
        \draw (0,0) -- (1.6,-1.6);
        \draw (2,0) -- (1,-1);
        \draw [thick, loosely dotted] (1.6,-1.6) -- (2.4,-2.4);
        \node [below] at (2, -2.6) {\scriptsize $\vb{x}$};
        \draw (6,0) -- (3,-3);
        \draw (8,0) -- (11,-3);
        \draw (12,0) -- (13,-1);
        \draw (14,0) -- (12.4,-1.6);
        \draw (2.4,-2.4) -- (3,-3) -- (7,-4) -- (7,-5);
        \draw [thick, loosely dotted] (12.4,-1.6) -- (11.6,-2.4);
        \node [below] at (12, -2.6) {\scriptsize $\vb{y}$};
        \draw (11.6,-2.4) -- (11,-3) -- (7,-4);
        \node [right] at (7,-4.5) {\scriptsize $J$};
        \node at (4.6,-4) {\scriptsize $\alpha$};

        \begin{scope}[shift={(0,-4.5)}, yscale=-1, shift={(0,4.5)}]
        \draw (0,0) -- (1.6,-1.6);
        \draw (2,0) -- (1,-1);
        \draw [thick, loosely dotted] (1.6,-1.6) -- (2.4,-2.4);
        \node [above] at (2, -2.6) {\scriptsize $\vb{x}'$};
        \draw (6,0) -- (3,-3);
        \draw (8,0) -- (11,-3);
        \draw (12,0) -- (13,-1);
        \draw (14,0) -- (12.4,-1.6);
        \draw (2.4,-2.4) -- (3,-3) -- (7,-4);
        \draw [thick, loosely dotted] (12.4,-1.6) -- (11.6,-2.4);
        \node [above] at (12, -2.6) {\scriptsize $\vb{y}'$};
        \draw (11.6,-2.4) -- (11,-3) -- (7,-4);
        \node [right] at (7,-4.5) {\scriptsize $J$};
        \node at (4.6,-4) {\scriptsize $\alpha'$};
        \end{scope}
    \end{tikzpicture}\,,
\end{equation}
where $\Psi_{\vb{x},\alpha,\vb{y}}\in\mathbb{C}$ are the normalized coefficients of the state in the fusion basis labeled by intermediate charges. Foreshadowing the partial trace, we define $p_\alpha=\sum_{\vb{x},\vb{y}}\abs{\Psi_{\vb{x},\alpha,\vb{y}}}^2$ and $\psi_{\vb{x},\alpha,\vb{y}}=\frac{1}{\sqrt{p_\alpha}}\Psi_{\vb{x},\alpha,\vb{y}}$, satisfying $\sum_\alpha p_\alpha=1$ and $\sum_{\vb{x},\vb{y}}\abs{\psi_{\vb{x},\alpha,\vb{y}}}^2=1$. Performing the partial trace gives the reduced anyonic density matrix
\begin{widetext}
\begin{equation}
\label{eq:reduced_density_matrix}
    \tilde{\rho}_A=\sum_{\alpha,\alpha'}\sqrt{p_\alpha p_{\alpha'}}\sum_{\substack{\vb{x},\vb{y} \\ \vb{x}',\vb{y}'}}\frac{\psi_{\vb{x},\alpha,\vb{y}}\psi^*_{\vb{x}',\alpha',\vb{y}'}}{\sqrt{d_J d_{\mathfrak{j}}^L}} \hspace{-0.5cm}\begin{tikzpicture}[baseline={([yshift=-0.5ex]current bounding box.center)},x=\cell cm,y=\cell cm]
        \draw (0,0) -- (1.6,-1.6);
        \draw (2,0) -- (1,-1);
        \draw [thick, loosely dotted] (1.6,-1.6) -- (2.4,-2.4);
        \node [below] at (2, -2.6) {\scriptsize $\vb{x}$};
        \draw (6,0) -- (3,-3);
        \draw (8,0) -- (11,-3);
        \draw (12,0) -- (13,-1);
        \draw (14,0) -- (12.4,-1.6);
        \draw (2.4,-2.4) -- (3,-3) -- (7,-4) -- (7,-5);
        \draw [thick, loosely dotted] (12.4,-1.6) -- (11.6,-2.4);
        \node [below] at (12, -2.6) {\scriptsize $\vb{y}$};
        \draw (11.6,-2.4) -- (11,-3) -- (7,-4);
        \node [right] at (7,-4.5) {\scriptsize $J$};
        \node at (4.6,-4) {\scriptsize $\alpha$};

        \draw (14,0) -- (15,-1) -- (15,-4.5);
        \draw (12,0) -- (14,2) -- (16,0) -- (16,-4.5);
        \draw (8,0) -- (9.2, 1.2);
        \draw [thick, loosely dotted] (9.2,1.2) -- (10,2);

        \begin{scope}[shift={(0,-4.5)}, yscale=-1, shift={(0,4.5)}]
        \draw (0,0) -- (1.6,-1.6);
        \draw (2,0) -- (1,-1);
        \draw [thick, loosely dotted] (1.6,-1.6) -- (2.4,-2.4);
        \node [above] at (2, -2.6) {\scriptsize $\vb{x}'$};
        \draw (6,0) -- (3,-3);
        \draw (8,0) -- (11,-3);
        \draw (12,0) -- (13,-1);
        \draw (14,0) -- (12.4,-1.6);
        \draw (2.4,-2.4) -- (3,-3) -- (7,-4);
        \draw [thick, loosely dotted] (12.4,-1.6) -- (11.6,-2.4);
        \node [above] at (12, -2.6) {\scriptsize $\vb{y}'$};
        \draw (11.6,-2.4) -- (11,-3) -- (7,-4);
        \node [right] at (7,-4.5) {\scriptsize $J$};
        \node at (4.6,-4) {\scriptsize $\alpha'$};

        \draw (14,0) -- (15,-1) -- (15,-4.5);
        \draw (12,0) -- (14,2) -- (16,0) -- (16,-4.5);
        \draw (8,0) -- (9.2, 1.2);
        \draw [thick, loosely dotted] (9.2,1.2) -- (10,2);
        \end{scope}
    \end{tikzpicture}=\sum_{\alpha}p_\alpha\sum_{\vb{x},\vb{x}',\vb{y}}  \frac{\psi_{\vb{x},\alpha,\vb{y}}\psi^*_{\vb{x}',\alpha,\vb{y}}}{\sqrt{d_\alpha d_\mathfrak{j}^{L_A}}}\hspace{-0.5cm}\begin{tikzpicture}[baseline={([yshift=-0.5ex]current bounding box.center)},x=\cell cm,y=\cell cm]
        \draw (0,0) -- (1.6,-1.6);
        \draw (2,0) -- (1,-1);
        \draw [thick, loosely dotted] (1.6,-1.6) -- (2.4,-2.4);
        \draw (6,0) -- (3,-3);
        \draw (2.4,-2.4) -- (3,-3) -- (3,-4);
        \node [right] at (3,-4) {\scriptsize $\alpha$};
        \node [below] at (2, -2.4) {\scriptsize $\vb{x}$};

        \begin{scope}[shift={(0,-4)}, yscale=-1, shift={(0,4)}]
        \draw (0,0) -- (1.6,-1.6);
        \draw (2,0) -- (1,-1);
        \draw [thick, loosely dotted] (1.6,-1.6) -- (2.4,-2.4);
        \draw (6,0) -- (3,-3);
        \draw (2.4,-2.4) -- (3,-3) -- (3,-4);
        \node [right] at (3,-4) {\scriptsize $\alpha$};
        \node [above] at (2, -2.4) {\scriptsize $\vb{x}'$};
        \end{scope}
    \end{tikzpicture}\,.
\end{equation}
\end{widetext}
We identify the RHS as the decomposition of $\tilde{\rho}_A$ into the $\alpha$ charge sectors, $\tilde{\rho}_A=\bigoplus_\alpha p_\alpha \tilde{\rho}_{A\alpha}$. It is clear from this that $p_\alpha$ has the interpretation of the probability to be in the $\alpha$ sector. The coefficients $\psi_{\vb{x},\alpha,\vb{y}}$ define a matrix $[R_\alpha]_{\vb{x},\vb{x}'}=\sum_{\vb{y}}\psi_{\vb{x},\alpha,\vb{y}}\psi^*_{\vb{x'},\alpha,\vb{y}}$ with unit trace. Since each $\tilde{\rho}_{A\alpha}$ are mutually orthogonal, the AEE splits into two pieces $\tilde{S}_A=H(\{p_\alpha\})+\sum_\alpha p_\alpha \tilde{S}(\tilde{\rho}_{A\alpha})$, a classical Shannon entropy $H$ of the probability distribution, and a weighted average of AEEs in each sector.

It remains to calculate $\tilde{S}(\tilde{\rho}_{A\alpha})=-\partial_s \tilde{\tr}\,(\tilde{\rho}_{A,\alpha})^s\,\big\rvert_{s=1}$ via the replica trick. Stacking diagrams $s$ times and evaluating,
\begin{equation}
    (\tilde{\rho}_{A\alpha})^s=\frac{1}{d_\alpha^s}\sum_{\vb{x},\vb{x}'}[R^s_\alpha]_{\vb{x},\vb{x}'}\sqrt{\frac{d_\alpha}{d_\mathfrak{j}^{L_A}}}\begin{tikzpicture}[baseline={([yshift=-0.5ex]current bounding box.center)},x=\cell cm,y=\cell cm]
        \draw (0,0) -- (1.6,-1.6);
        \draw (2,0) -- (1,-1);
        \draw [thick, loosely dotted] (1.6,-1.6) -- (2.4,-2.4);
        \draw (6,0) -- (3,-3);
        \draw (2.4,-2.4) -- (3,-3) -- (3,-4);
        \node [right] at (3,-4) {\scriptsize $\alpha$};
        \node [below] at (2, -2.4) {\scriptsize $\vb{x}$};

        \begin{scope}[shift={(0,-4)}, yscale=-1, shift={(0,4)}]
        \draw (0,0) -- (1.6,-1.6);
        \draw (2,0) -- (1,-1);
        \draw [thick, loosely dotted] (1.6,-1.6) -- (2.4,-2.4);
        \draw (6,0) -- (3,-3);
        \draw (2.4,-2.4) -- (3,-3) -- (3,-4);
        \node [right] at (3,-4) {\scriptsize $\alpha$};
        \node [above] at (2, -2.4) {\scriptsize $\vb{x}'$};
        \end{scope}
    \end{tikzpicture}\,.
\end{equation}
The quantum trace of the diagram gives
\begin{equation}
    \tilde{\tr}\,(\tilde{\rho}_{A\alpha})^s=\frac{1}{d_\alpha^{s-1}}\tr R^s_\alpha\,,
\end{equation}
yielding the total formula
\begin{equation}
\label{eq:state_AEE}
    \tilde{S}_A=H(\{p_\alpha\})+\sum_\alpha p_\alpha(H(\{\lambda_{\alpha,i}\})+\log(d_\alpha))\,,
\end{equation}
where $\lambda_{\alpha,i}$ are the eigenvalues of the matrix $R_\alpha$. Compared to the usual von Neumann entropy, Eq.~\eqref{eq:state_AEE} picks up a topological contribution $\log(d_\alpha)$ which quantifies accessible entanglement in the limit of many copies.

For computing the average AEE, we follow the steps in Ref.~\cite{BD}\@. When $\tilde{\rho}$ is drawn at random, the weights $p_\alpha$ and matrices $R_\alpha$ are themselves random variables. The Haar-uniform measure on $\mathcal{H}^J$ induces a Dirichlet distribution on $p_\alpha$, parametrized by the dimensions of the corresponding $\alpha$ sectors, and a fixed-trace Wishart--Laguerre distribution on each $R_\alpha$. Performing the averaging,
\begin{align}
    \label{eq:exact_AEE}
    \langle\tilde{S}_A\rangle_J&=\sum_\alpha \varrho_\alpha \varphi_\alpha\,, \quad \text{where} \quad \varrho_\alpha=\frac{m_\alpha n_\alpha}{D_J}\,, \nonumber \\
    \varphi_\alpha&=\Psi(D_J+1)-\Psi(\max(m_\alpha,n_\alpha)+1) \nonumber \\
    &\phantom{={}}-\min\left(\frac{m_\alpha-1}{2n_\alpha},\frac{n_\alpha-1}{2m_\alpha}\right)+\log(d_\alpha)\,.
\end{align}
Here
\begin{align}
\label{eq:dimensions_in_AEE}
    m_\alpha=D_\alpha(L_A)\,, \quad n_\alpha=\sum_{\beta\mid N^J_{\alpha\beta}=1}D_\beta(L_B)\,, \nonumber \\
    D_J=D_J(L) \quad \text{with} \quad D_J=\sum_\alpha m_\alpha n_\alpha
\end{align} are Hilbert space dimensions adapted to the decomposition $\bigoplus_{\alpha} \left(\mathcal{H}^\alpha_A\otimes\left(\bigoplus_\beta\mathcal{H}^\beta_B\otimes V_{\alpha\beta}^J\right)\right)$. It is useful to note that $\varrho_\alpha$ defines a probability distribution, and $\langle\tilde{S}_A\rangle_J$ is an average of the observable $\varphi_\alpha$ with respect to it. Finally, $\Psi$ denotes the digamma function $\Psi(x)=\partial_x \log(\Gamma(x))$.

This derivation provides the ingredients for calculating the average anyonic mutual information $\langle\tilde{S}(A:B)\rangle_J=\langle\tilde{S}_A\rangle_J+\langle\tilde{S}_B\rangle_J-\log(d_J)$, where we used the fact that $\tilde{S}_{AB}=\log(d_J)$ for a state supported in a fixed $J$ sector. $\langle \tilde{S}_B\rangle_J$ can be calculated by swapping the $A$ and $B$ subsystems and using the appropriate dimensions in Eq.~\eqref{eq:dimensions_in_AEE}.

\subsection{Large $L$ behavior and examples}
\label{sec:large-L}
Eq.~\eqref{eq:exact_AEE} is an exact formula for the AEE of a bipartite state with total charge $J$. We are interested in its large $L$ expansion to understand its leading and subleading behavior in the thermodynamic limit. The general strategy is to note that Eq.~\eqref{eq:exact_AEE} depends only on the fusion space dimensions, which for UMTCs, can be calculated exactly by the Verlinde formula
\begin{equation}
\label{eq:Verlinde_formula}
    N^c_{ab}=\sum_{j\in\mathcal{S}} \frac{S_{aj}S_{bj}S_{cj}^*}{S_{0j}}\,,
\end{equation}
where $S$ is the modular $S$-matrix of the theory. The Verlinde formula tells us that the $S$-matrix diagonalizes the fusion matrix $N_a$, in the sense that $(N_a)_{bc}=\sum_{j\in\mathcal{S}} S_{bj}\frac{S_{aj}}{S_{0j}}S_{jc}^*$. We also have that the fusion space dimension is given by
\begin{align}
    \dim V^J_{a_1\dots a_L}&=\sum_{x_2,x_3\dots,x_{L-1}}N^{x_2}_{a_1a_2}N^{x_3}_{x_2a_3}\cdots N^J_{x_{L-1}a_L} \nonumber \\
    &=\left(\prod_{i=1}^L N_{a_i}\right)_{0J}\,,\label{eq:fusion_multiplicity_matrix}
\end{align}
a matrix element of a product of fusion matrices. It is easy then to see that 
\begin{equation}
    \dim V^J_{a_1\dots a_L}=\sum_{j\in\mathcal{S}}S_{0j}S_{Jj}^*\prod_{i=1}^L\frac{S_{a_ij}}{S_{0j}}\,.
\end{equation}
For our physical setup where all anyons $a_i=\mathfrak{j}$ are identical,
\begin{equation}
\label{eq:Verlinde_dimension}
    D_J(L)=\sum_{j\in\mathcal{S}}S_{0j}S_{Jj}^*\left(\frac{S_{\mathfrak{j}j}}{S_{0j}}\right)^L\,.
\end{equation}
We now specify to the explicit anyon models introduced in Section~\ref{sec:anyon_examples}.

\subsubsection{Abelian anyon models}
For abelian anyons, the fusion spaces are all one-dimensional when they exist, so $\dim V^J_{a_1,\dots,a_L}=\delta_{J,a_1\cdots a_L}$. That means, given a state with total charge $J$, for any bipartition of the system, the charge $\alpha$ is fixed and $D_J=m_\alpha=n_\alpha=1$. Also, all abelian anyons have $d_\alpha=1$. The anyonic state must factor across the bipartition and, as verified by the formulae~\eqref{eq:exact_AEE}, $\langle\tilde{S}_A\rangle_J=0$.

\subsubsection{$\SU(2)_k$ anyons}
For $\SU(2)_k$, one has the modular $S$-matrix~\cite[Section~5.4]{Bonderson_thesis}
\begin{equation}
    S_{ab}=\sqrt{\frac{2}{k+2}}\sin\left(\frac{(2a+1)(2b+1)\pi}{k+2}\right)\,.
\end{equation}
Write $\lambda_j=S_{\mathfrak{j} j}/S_{0j}$, which are eigenvalues of $N_\mathfrak{j}$. Under the change of indices $j\mapsto k/2-j$, one shows that $S_{Jj}\mapsto (-1)^{2J}S_{Jj}$. Therefore, we can rewrite the sum~\eqref{eq:Verlinde_dimension} as
\begin{align}
\label{eq:dimension_Verlinde}
    D_J(L)&=(1+(-1)^{2\mathfrak{j}L+2J})\sum_{j\in\mathcal{S}'} S_{0j}S_{Jj}\lambda_j^L \nonumber \\
    &\phantom{={}}+\delta_{k\text{ even}}S_{0\frac{k}{4}}S_{J\frac{k}{4}}\lambda_\frac{k}{4}^L\,,
\end{align}
where $\mathcal{S}'\subset\mathcal{S}$ contains only the first half anyon charges (excluding the fixed-point term when $k$ is even, which we separated out). The prefactor is a consistency relation which ensures that the dimension vanishes for invalid configurations, \eg an odd chain of spin-$1/2$ anyons cannot have integer total spin. The sum~\eqref{eq:dimension_Verlinde} is always dominated by $\lambda_0=d_\mathfrak{j}$, which has the largest modulus. Thus,
\begin{equation}
    D_J(L)=(1+(-1)^{2\mathfrak{j}L+2J})S_{00}S_{J0}d_\mathfrak{j}^L(1+o(1))\,.
\end{equation}
Analogously for the other dimensions in Eq.~\eqref{eq:dimensions_in_AEE},
\begin{align}
    \begin{split}
        m_\alpha&=(1+(-1)^{2\mathfrak{j}L_A+2\alpha})S_{00}S_{\alpha0}d_\mathfrak{j}^{L_A}(1+o(1))\,, \\
        n_\alpha&=(1+(-1)^{2\mathfrak{j}L_B+2(\alpha+J)})S_{\alpha0}S_{J0}d_\mathfrak{j}^{L_B}(1+o(1))\,.
    \end{split}
\end{align}
Note that these approximations fail in the regime $k^2\gg L$, where many $\lambda_j$ cluster near $\lambda_0$. In this case, the sum is approximated by an integral and is equivalent to the method of Weyl character formulas presented in Refs.~\cite{EESU(2)1,SREESU(2)}.

Without loss of generality, we drop the prefactors and assume we have a valid configuration of spins. Define the subsystem size ratio as $f=L_A/L$---we analyze the cases $f\leq 1/2$ and $f>1/2$ separately. 

Before we take the large $L$ limit, we make a comment on the non-analyticity of $\varphi_\alpha$ at finite $L$. It occurs at the point $m_\alpha=n_\alpha$, which can be solved for $f$ to give $f_*=1/2+\log(d_J)/(2L\log(d_\mathfrak{j}))+o(1/L)$. The discontinuity will always lie above $f=1/2$, but as $L$ becomes large, the discontinuity will approach $f=1/2$.

Substituting these expressions into the observable $\varphi_\alpha$ and using the expansion $\Psi(x)=\log(x)+o(1)$, we simplify to
\begin{widetext}
    \begin{equation}
    \varphi_\alpha=\begin{cases}
        fL\log(d_\mathfrak{j})+\log\left(\frac{S_{00}}{S_{\alpha0}}\right)+\log(d_\alpha)-\frac{S_{00}}{2S_{J0}}\delta_{f,\frac{1}{2}}+o(1)\,, & f\leq \frac{1}{2} \\
        (1-f)L\log(d_\mathfrak{j})+\log\left(\frac{S_{J0}}{S_{\alpha0}}\right)+\log(d_\alpha)+o(1)\,, & f>\frac{1}{2}
    \end{cases}\,.
    \end{equation}
\end{widetext}
In the large $L$ expansion, the discontinuity of the observable is exactly at $f=1/2$, which is exponentially suppressed away from half-bipartition. The surprising finding is that in both cases, the quantum dimension $d_\alpha$ is related to the $S$-matrix $d_\alpha=S_{\alpha 0}/S_{00}$ precisely as to make $\varphi_\alpha$ independent of $\alpha$. Recalling that $\varrho_\alpha$ is just a probability distribution, that means the average AEE is simply
\begin{equation}
\label{eq:SU(2)_k_AEE}
    \langle \tilde{S}_A\rangle_J=\begin{cases}
    fL\log(d_\mathfrak{j})-\frac{1}{2d_J}\delta_{f,\frac{1}{2}}+o(1)\,, & f\leq \frac{1}{2} \\
    (1-f)L\log(d_\mathfrak{j})+\log(d_J)+o(1)\,, & f>\frac{1}{2}
    \end{cases}\,.
\end{equation}

A few remarks are in order. Firstly, the leading $O(L)$ behavior is independent of the total charge sector $J$ of the global state. To get $J$ dependence, one would need a non-vanishing spin density in the thermodynamic limit. However, that is not possible within the $\SU(2)_k$ framework for a fixed level $k$, where there are finitely many spins. The coefficient of this term is predicted by the effective local Hilbert space dimension of each site, which follows from a density of states argument.

While the $O(L)$ term is symmetric under subsystem exchange $f\leftrightarrow 1-f$, the $O(1)$ term is not when $J$ is non-abelian.  The same phenomenon was noted already in Ref.~\cite{SREESU(2)} in the context of $\SU(2)$ symmetry-resolved entanglement, and is attributed to an effect of superselection, following from the fact that the commutant of the observable algebra in $B$ is not equal to the observable algebra in $A$, \ie $\mathcal{M}_A\neq (\mathcal{M}_B)'$. In addition, the $\log(d_J)$ jump as one crosses from $f<1/2$ to $f>1/2$ is indicative of a non-uniform convergence, which can only be resolved in the double scaling limit when $\abs{f-1/2}=O(1/L)$ \cite{typicalEEreview,EEU(1)2}, which we pursue in the next section.

The expression for $\langle \tilde{S}_A\rangle$, except for the extra factor of $1/d_J$ in the Page correction, greatly resembles the original Page curve (lack of $O(1)$ terms). This is surprising, because that calculation was performed in a tensor product space in the absence of symmetries. All preceding works \cite{EEU(1)1,EEU(1)2,EESU(2)1,EESU(2)2,SREESU(2)} suggested, that even in the $J=0$ sector, there would at the very least be $O(1)$ corrections in the presence of symmetries. In many cases, these $O(1)$ terms were $f$-dependent and appeared in the combination $(f+\log(1-f))/2$. These features are lacking in our formulae. The cause of this is two-fold: first, the finite structure of the fusion category forces Hilbert space dimensions to scale only exponentially $\sim \rme^{\beta L}$, whereas in $\SU(2)$ systems, generally one has a scaling behavior $\sim L^{-\alpha} \rme^{\beta L}$. The extra factors in front of the exponential require non-trivial treatment and give rise to the non-trivial $f$-dependent terms in these cases. Second, the choice of entanglement measure that we used, the AEE, which uses the quantum trace, behaves well under tensor products of operators and respects charge superselection. We study the entanglement entropy if the standard von Neumann entropy were used instead in Appendix~\ref{app:vN_entropy}, where there are interesting $O(1)$ contributions.

\subsubsection{Fibonacci anyons}
In Section~\ref{sec:eigenstate_AEE}, we study the AEE of eigenstates of Hamiltonians in the Fibonacci theory. The analytical results presented in this section will serve as a benchmark for our numerics. For the sake of umambiguity, we report the Haar-random AEE using the Fibonacci  quantum dimension $d_\tau=\phi=(1+\sqrt{5})/2$,
\begin{equation}
\label{eq:Fib_AEE}
    \langle \tilde{S}_A\rangle_J=\begin{cases}
    fL\log(\phi)-\frac{1}{2d_J}\delta_{f,\frac{1}{2}}+o(1)\,, & f\leq \frac{1}{2} \\
    (1-f)L\log(\phi)+\log(d_J)+o(1)\,, & f>\frac{1}{2}
    \end{cases}\,.
\end{equation}

\subsection{Resolving non-analyticities at $f=1/2$}
The $L\to\infty$ limit of the entanglement entropy \eqref{eq:SU(2)_k_AEE} is non-uniform because the pointwise limit is not continuous. Its origin derives from two non-analytic pieces: a maximum of the subsystem dimensions $\max(m_\alpha,n_\alpha)$, and a ratio of dimensions $\min(m_\alpha/n_\alpha,n_\alpha/m_\alpha)$. As we saw, the first contributes a jump of $\log(d_J)$ across $f=1/2$, and the second contributes a Kronecker delta at the same point. Both terms can be resolved to be a continuous function of $f$, if, in the thermodynamic limit, we simultaneously allow $\abs{f-1/2}$ to scale in a certain way with the system size.

We can resolve the Kronecker delta via a double scaling limit, where one takes the limit of a sequence of functions and simultaneously zooms into the interval around a point where this sequence is non-uniformly convergent. A prime example is the sequence $f_L(x)=x^L$ on the interval $x\in[0,1]$ with the point-wise limit ${\lim_{L\to\infty} f_L=\delta_{x,1}}$, which is non-uniformly convergent at $x=1$. We can resolve this Kronecker delta by studying the limit $f=f_L(1+\Lambda/L^s)$ for $\Lambda \in(-\infty,0]$ and $s>0$. Here, $s$ describes the scale of zooming into the interval around $x=1$. We find
\begin{align}
    \hspace{-2mm}\lim_{L\to\infty}f_L(1+\tfrac{\Lambda}{L^s})\!=\!\lim_{L\to\infty}\left(1+\frac{\Lambda}{L^s}\right)^L\!=\!\begin{cases}
    1\,, & s>1\\
    \rme^{\Lambda}\,, & s=1\\
    0\,, & s<1
    \end{cases},
\end{align}
which thus demonstrates that our sequence of functions ``looks'' like the exponential function on a scale inversely proportional to $L$ around $x=1$. We see that $\rme^{\Lambda}$ continuously interpolates between $f(x<1)=0$ for $\Lambda\to-\infty$ and $f(1)=1$ for $\Lambda=0$. We will now employ the same strategy to resolve Kronecker delta in Eq.~\eqref{eq:Fib_AEE}.

We revisit the ratio $m_\alpha/n_\alpha$ in Eq.~\eqref{eq:exact_AEE} and note that for non-trivial behavior, we require $(2f-1)L=O(1)$, or equivalently, that $f=1/2+\Lambda/(2L^s)$, where $s=1$ and $\Lambda$ is some $O(1)$ constant, \ie independent of $L$ when we take the limit. The crossover point, \ie where $m_\alpha=n_\alpha$, occurs at a finite non-zero value $\Lambda_c=\log(d_J)/\log(d_\mathfrak{j})$. We analyze both the difference of digammas and ratio of dimensions to determine the behavior of the average AEE in the crossover regime, and also the behaviours outside. If $s<1$, the approach is too slow and the terms pick a branch, yet if $s>1$, the terms always takes its value at the $f=1/2$ point. Up to constant order, we have
\begin{widetext}
\begin{equation}
\label{eq:resolved_delta}
\langle \tilde{S}_A\rangle=
\begin{cases}
    L_A \log(d_\mathfrak{j})+o(1)\,, & s<1 \text{ and } \Lambda\leq0\\
    L_B \log(d_\mathfrak{j})+\log(d_J)+o(1)\,, & s<1 \text{ and } \Lambda>0\\
    L_A\log(d_\mathfrak{j})-\max(0,\eta)-\frac{1}{2}\exp[-\abs{\eta}]+o(1)\,, & s=1\,, \eta=\Lambda\log(d_\mathfrak{j})-\log(d_J) \\
    L_A\log(d_\mathfrak{j})-\frac{1}{2d_J}+o(1)\,, & s>1 \\
\end{cases}\,.
\end{equation}
\end{widetext}
Let us briefly examine the behavior of the constant term in the $s=1$ regime. It is exponentially suppressed when $\eta<0$, and at exactly $\Lambda=0$, it gives rise to the Page correction $-1/(2d_J)$. When $\eta>0$, the term scales linearly in $\eta$ and contributes the $\log(d_J)$ jump across the half-bipartition.

\subsection{Variance}\label{sec:variance}
We report the variance of the AEE $(\Delta\tilde{S}_A)^2_J = \langle \tilde{S}_A^2\rangle_J - \langle \tilde{S}_A\rangle^2_J$ and its leading order. The variance can be calculated as
\begin{equation}
\label{eq:exact_variance}
    (\Delta\tilde{S}_A)^2_J=\frac{1}{D_J+1}\left(\sum_\alpha \varrho_\alpha(\varphi_\alpha^2+\chi_\alpha)-\langle \tilde{S}_A\rangle^2_J\right)\,,
\end{equation}
with $\varrho_\alpha$ and $\varphi_\alpha$ defined as in Eq.~\eqref{eq:exact_AEE} and

\begin{align}
    \chi_\alpha={}&(m_\alpha+n_\alpha)\Psi'(\max(m_\alpha,n_\alpha)+1)\nonumber\\
    &-(D_J+1)\Psi'(D_J+1) \nonumber \\
    &-\min\!\left(\frac{(m_\alpha-1)(m_\alpha+2n_\alpha-1)}{4n_\alpha^2},\right.\nonumber\\
    &\qquad\qquad\left.\frac{(n_\alpha-1)(n_\alpha+2m_\alpha-1)}{4m_\alpha^2}\right)\,.
\end{align}
The leading-order behavior comes from expanding $\Psi'(x)=1/x+O(1/x^2)$ and we find
\begin{equation}
    \chi_\alpha=\begin{cases}
        \frac{1}{2d_J}d_\mathfrak{j}^{-(1-2f)L}(1+o(1))\,, & f<\frac{1}{2} \\
        \frac{1}{2d_J}-\frac{1}{4d_J^2}+o(1)\,, &f=\frac{1}{2} \\
        \frac{1}{2d_J}d_\mathfrak{j}^{-(2f-1)L}(1+o(1))\,, & f>\frac{1}{2}
    \end{cases}\,.
\end{equation}
Putting this altogether in Eq.~\eqref{eq:exact_variance}, we obtain a simple asymptotic formula for the variance of the AEE:
\begin{equation}
\label{eq:SU(2)_k_variance}
    (\Delta\tilde{S}_A)^2_J\propto\begin{cases}
        d_\mathfrak{j}^{-2(1-f)L}(1+o(1))\,, & f<\frac{1}{2} \\
        d_\mathfrak{j}^{-L}(1+o(1))\,, & f=\frac{1}{2} \\
        d_\mathfrak{j}^{-2fL}(1+o(1))\,, & f>\frac{1}{2} \\
    \end{cases}\,.
\end{equation}
The upshot is that the variance always vanishes exponentially, so that the average AEE formula~\eqref{eq:SU(2)_k_AEE} describes the typical entanglement---the chance of selecting a Haar-random state with AEE not described by~\eqref{eq:SU(2)_k_AEE} is vanishingly small.

\subsection{Anyons on a torus}

The previous subsections considered anyons on a disk, but Eqs.~\eqref{eq:state_AEE} and \eqref{eq:exact_AEE} also describe the bipartite entanglement of anyons on a torus, provided one inserts the appropriate fusion space dimensions. The derivation is paralleled by starting with a global pure state $\ket{\Psi}\in V^x_{\overline{x} \mathfrak{j}^{\times L}}\otimes V_{x\overline{x}}^0$ (see Appendix~\ref{app:PBC}), and tracing out the boundaries $x,\overline{x}$ and anyons in subsystem $B$, which leaves a reduced density matrix of the same form as Eq.~\eqref{eq:reduced_density_matrix}. The interpretation, however, is slightly different. Because the torus is a manifold without boundary, the superselection label $x$ is not a total topological charge, but rather a charge threading a non-contractible cycle of the torus. More details can be found in Refs.~\cite{Bonderson_entanglement, Pfeifer_PBC}\@. The total charge of the system is always $J=0$ (abelian), so that the Page curve is recovered exactly without any topological corrections. For the explicit fusion space dimensions for anyons on a torus, we refer to Appendix~\ref{app:PBC}\@.

\section{Eigenstate entanglement of quantum-chaotic Hamiltonians}
\label{sec:eigenstate_AEE}
So far, we discussed the Haar-random average AEE in Hilbert space, with no reference to dynamics. We shift our focus now to eigenstates of quantum-chaotic anyonic Hamiltonians, and conjecture that the average midpspectrum eigenstate AEE coincides with the Haar-random average. This supports beliefs, stemming originally from the eigenstate thermalization hypothesis \cite{ETH1,ETH2,ETHanyons} and articulated more recently in~\cite{XXZEE,typicalEEreview,EESU(2)1,EEU(1)2}, that such eigenstates appear thermal and should have extensive entropy and resemble random states.

\subsection{Golden chain on a disk}\label{sec:OBC}

We study the golden chain Hamiltonian \cite{golden_chain,Fibonacci_intro} with nearest-neighbor (NN) and next-to-nearest-neighbor (NNN) terms. This model describes local antiferromagnetic interactions between $L$ Fibonacci anyons. Recall that the Fibonacci theory only has two anyons $\mathcal{S}=\{0,\tau\}$, corresponding to the integer charge sector $\{0,1\}$ of $\SU(2)_3$, \ie $\mathfrak{j}=1$. The Hamiltonian with open boundary conditions (OBC) is given by
\begin{equation}
\label{eq:golden_chain_OBC}
    H^{\mathrm{OBC}}=-\sum_{i=1}^{L-1}\Pi^{(i,i+1)}_0-\lambda \sum_{i=1}^{L-2}\Pi^{(i,i+2)}_0\,,
\end{equation}
where $\Pi^{(i,i+1)}_0$ projects the fusion channel of anyons $a_i$ and $a_{i+1}$ onto the trivial charge $0$. The NNN term deserves some further explanation. Anyons must be adjacent to define a fusion channel, so first anyons $a_i$, $a_{i+1}$ are braided to bring $a_i$ and $a_{i+2}$ adjacent; see Appendix~\ref{app:anyon_diagrams} for details. Then one projects using $\Pi^{(i+1,i+2)}_0$ and braids back in the opposite orientation to ensure there are no knots created. In performing the braids, an orientation (clockwise or anti-clockwise) was chosen. We choose to symmetrize the final term, or equivalently, take the real part of the matrix element. Diagrammatically,
\begin{equation}
\label{eq:NNN_term}
    \begin{tikzpicture}[baseline={([yshift=-0.5ex]current bounding box.center)},x=\cell cm,y=\cell cm]]
        \draw (-1.5,0) -- (4.5,0);
        \draw (0,0) -- (0,1.5);
        \draw (1.5,0) -- (1.5,1.2);
        \draw (3,0) -- (3,1.5);
        \draw[dashed] (1.5,1.2) arc (-90:90:0.8);
        \draw (0,2.5) -- (0,4) node [above] {\scriptsize $\tau$};
        \draw (1.5,2.8) -- (1.5,4) node [above] {\scriptsize $\tau$};
        \draw (3,2.5) -- (3,4) node [above] {\scriptsize $\tau$};
        \draw (-0.5,2.5) rectangle (3.5,1.5);
        \node at (1.5,2) {\scriptsize $\Pi_0$};
        \node[below] at (-0.75,0) {\scriptsize $x_{i-1}$};
        \node[below] at (0.75,0) {\scriptsize $x_i$};
        \node[below] at (2.25,0) {\scriptsize $x_{i+1}$};
        \node[below] at (3.75,0) {\scriptsize $x_{i+2}$};
    \end{tikzpicture}\hspace{-4mm}=
    \frac{1}{2}\left(\begin{tikzpicture}[baseline={([yshift=-0.5ex]current bounding box.center)},x=\cell cm,y=\cell cm]]
        \draw (-1.5,0) -- (4.5,0);
        \begin{knot}[clip width=3]
            \strand (0,0) .. controls +(0,1) and +(0,-1) .. (1.5,1.5);
            \strand (1.5,0) .. controls +(0,1) and +(0,-1) .. (0,1.5) -- (0,1.5) -- (0,2.5);
        \end{knot}
        \draw (3,0) -- (3,1.5);
        \draw (3,2.5) -- (3,4) node [above] {\scriptsize $\tau$};
        \draw (1,2.5) rectangle (3.5,1.5);
        \node at (2.25, 2) {\scriptsize $\Pi_0$};
        \begin{knot}[clip width=3, flip crossing=1]
            \strand (0,2.5) .. controls +(0,1) and +(0,-1) .. (1.5,4) node [above] {\scriptsize $\tau$};
            \strand (1.5,2.5) .. controls +(0,1) and +(0,-1) .. (0,4) node [above] {\scriptsize $\tau$};
        \end{knot}
        \node[below] at (-0.75,0) {\scriptsize $x_{i-1}$};
        \node[below] at (0.75,0) {\scriptsize $x_i$};
        \node[below] at (2.25,0) {\scriptsize $x_{i+1}$};
        \node[below] at (3.75,0) {\scriptsize $x_{i+2}$};
    \end{tikzpicture}\hspace{-2mm}+\hspace{-2mm}\begin{tikzpicture}[baseline={([yshift=-0.5ex]current bounding box.center)},x=\cell cm,y=\cell cm]]
        \draw (-1.5,0) -- (4.5,0);
        \begin{knot}[clip width=3, flip crossing=1]
            \strand (0,0) .. controls +(0,1) and +(0,-1) .. (1.5,1.5);
            \strand (1.5,0) .. controls +(0,1) and +(0,-1) .. (0,1.5) -- (0,1.5) -- (0,2.5);
        \end{knot}
        \draw (3,0) -- (3,1.5);
        \draw (3,2.5) -- (3,4) node [above] {\scriptsize $\tau$};
        \draw (1,2.5) rectangle (3.5,1.5);
        \node at (2.25, 2) {\scriptsize $\Pi_0$};
        \begin{knot}[clip width=3]
            \strand (0,2.5) .. controls +(0,1) and +(0,-1) .. (1.5,4) node [above] {\scriptsize $\tau$};
            \strand (1.5,2.5) .. controls +(0,1) and +(0,-1) .. (0,4) node [above] {\scriptsize $\tau$};
        \end{knot}
        \node[below] at (-0.75,0) {\scriptsize $x_{i-1}$};
        \node[below] at (0.75,0) {\scriptsize $x_i$};
        \node[below] at (2.25,0) {\scriptsize $x_{i+1}$};
        \node[below] at (3.75,0) {\scriptsize $x_{i+2}$};
    \end{tikzpicture}\right).
\end{equation}
For more details and the explicit matrix elements, we refer to Ref.~\cite{Fibonacci_intro}\@. The action of the Hamiltonian leaves the total charge $J$ of the state invariant, \ie it is block-diagonal in $J$, and we can study each sector independently.

We remark that these Hamiltonians are anyonic generalizations of the spin-1/2 $\SU(2)$ Heisenberg model
\begin{equation}
    \label{eq:Heisenberg_Hamiltonian}
    H^{\SU(2)}=-\sum_{i=1}^{L-1}\vb{S}_i\cdot \vb{S}_{i+1}-\lambda \sum_{i=1}^{L-2}\vb{S}_i\cdot \vb{S}_{i+2}\,,
\end{equation}
because, up to a constant shift, each term in Eq.~\eqref{eq:Heisenberg_Hamiltonian} is a projector onto the singlet. Under the automorphism $j\mapsto k/2-j$, the Fibonacci anyon is truly the anyonic generalization of a spin-1/2 particle.

When $\lambda=0$, the golden chain~\eqref{eq:golden_chain_OBC} is known to be integrable and described by a conformal field theory with central charge $c=7/10$~\cite{golden_chain}\@. The NNN interaction term breaks integrability, with strength controlled by the coupling $\lambda\in\mathbb{R}$. Note that because of the choice of braiding convention, $\Pi^{(i,i+2)}_0$ does not commute with $\Pi^{(i+1,i+3)}_0$, and therefore the $\lambda\to\infty$ limit does not decouple into two integrable chains.

\begin{figure}[t]
    \includegraphics[width=\linewidth]{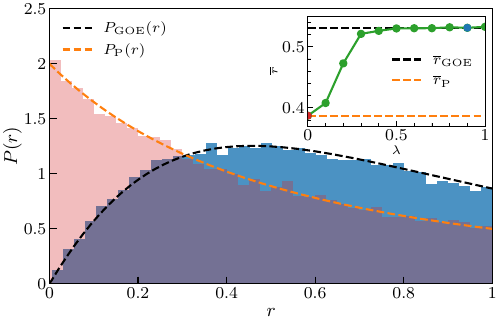}
    \caption{\justifying Distribution of level-spacing ratios $r_m$ of the $L=26$ golden chain Hamiltonian~\eqref{eq:golden_chain_OBC} with $\lambda=0.9\, [0]$ in the sector $J=0$, $\parity=+1$. The dashed chaotic GOE [integrable Poissonian] prediction $P_{\mathrm{GOE}}(r)=\frac{27}{4}(r+r^2)/(1+r+r^2)^{5/2}$ [$P_\mathrm{P}(r)=2/(1+r^2)$] is overlayed. Inset: the average level-spacing ratio $\overline{r}$ as a function of $\lambda$, compared to GOE $\overline{r}_{\mathrm{GOE}}\approx0.5307$ [Poissonian $\overline{r}_\mathrm{P}\approx0.3867$] predictions. The blue [red] dot corresponds to the value $\lambda=0.9\, [0]$ chosen in the main plot.}
    \label{fig:level_spacing_ratios}
\end{figure}

Exact diagonalization enables access to the full spectrum of the Hamiltonian \eqref{eq:golden_chain_OBC} for chain lengths up to $L=26$. We identify a quantum-chaotic regime at $\lambda=0.9$ for $J=0$ and we verify this using the level-spacing ratio~\cite{averageEEchaos2}
\begin{equation}
    r_m=\min\left(\frac{E_m-E_{m-1}}{E_{m+1}-E_m},\frac{E_{m+1}-E_m}{E_m-E_{m-1}}\right)\,,
\end{equation}
where the $E_m$ are the eigenenergies of the Hamiltonian.
The distribution is plotted in Figure~\ref{fig:level_spacing_ratios}, which shows good agreement with the prediction from the random Gaussian orthogonal ensemble (GOE). We also plot the integrable $\lambda=0$ regime, with good agreement to the Poisson distribution. Finding the level-spacing ratios involved resolving the additional reflection symmetry in the OBC chain, which splits the Hilbert space into $\pm1$ parity sectors.

\begin{figure}[t]
    \includegraphics[width=\linewidth]{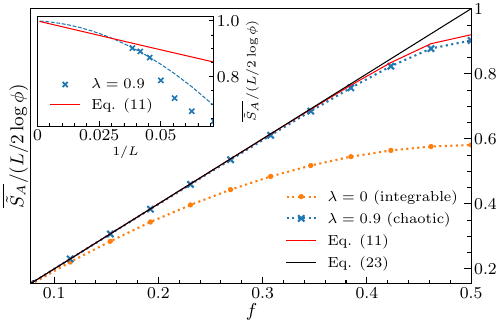}
    \caption{\justifying Normalized Page curve for the $L=26$ golden chain Hamiltonian~\eqref{eq:golden_chain_OBC} in the sector $J=0$, $\parity=+1$. We plot the AEE of integrable (orange) and quantum-chaotic (blue) eigenstates as a function of subsystem fraction $f=L_A/L$. We also show the exact analytical prediction~\eqref{eq:exact_AEE} (red) and its leading-order~\eqref{eq:Fib_AEE} (black). The Page curve for $J=0$ is symmetric across $f=0.5$. Inset: an attempt at finite-size scaling for the half-chain AEE of quantum-chaotic eigenstates. The dashed line is a fit $1+a/L+b/L^2$ to the three largest chain sizes, $L=22,24,26$. The poor scaling results might indicate that the chain lengths used are not yet in the asymptotic regime.}
    \label{fig:OBC_AEE}
\end{figure}

We compute the average AEE\footnote{We use angled-brackets $\langle\cdot\rangle$ to denote a Haar-random average and an overline $\overline{\cdot}$ to denote an arithmetic mean.} $\overline{\tilde{S}_A}$ of eigenstates of the golden chain Hamiltonian with the same parameters and sectors as in the study of the level statistics. To avoid the spectral edges, the average is performed only over the central 2000 eigenvectors. Figure~\ref{fig:OBC_AEE} shows this average as a function of $f=L_A/L$, reproducing the Page curve for anyonic systems. It matches well with the analytical prediction \eqref{eq:Fib_AEE}. In contrast, the average over eigenvectors in the integrable $\lambda=0$ regime deviates quickly from the chaotic behavior and clearly shows sub-maximal entanglement away from $f=0$. 

In previous studies \cite{averageEEchaos4,chaosKLdivergenceEE}, it was shown that the deviation between average eigenstate AEE and the Haar-random AEE occurs at constant $O(1)$ order. To verify this, we attempted a finite-size scaling analysis in the inset of Figure~\ref{fig:OBC_AEE}, where we obtained the average half-chain AEE of eigenstates for even system sizes from $L=14$ to $L=26$. For $L\leq20$, we observe a sizeable difference, consistent with strong finite-size effects coming from the small Hilbert space dimension $D_0(L)=F_{L-1}$, where $F_{L-1}$ are the Fibonacci numbers. The attempted phenomenological fit accounting for finite-size effects does not cleanly match the expected asymptotic scaling with system size. This could potentially indicate that the eigenstate EE of quantum chaotic Hamiltonians has a different correction term than the average Haar random analytics. However, we cannot make a fully conclusive statement here because the considered systems sizes might not yet be in the asymptotic regime, where the actual scaling becomes visible. Nevertheless, we can conclude that finite-size exact diagonalization of the quantum-chaotic golden chain Hamiltonian shows a clear approach towards the Haar-random prediction.

\begin{figure}[t]
    \includegraphics[width=\linewidth]{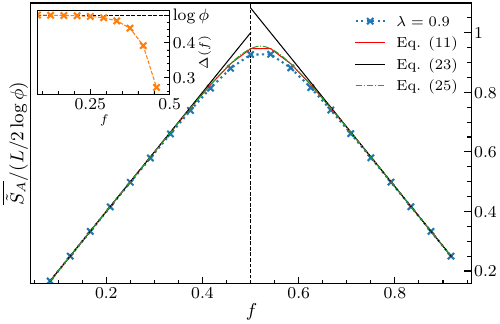}
    \caption{\justifying Asymmetric normalized Page curve for the $L=24$ golden chain Hamiltonian~\eqref{eq:golden_chain_OBC} with $\lambda=0.9$ in the sector $J=1$, $\parity=+1$, c.f.\@ Figure~\ref{fig:OBC_AEE}. We additionally plot the resolved curve Eq.~\eqref{eq:resolved_delta} at $s=1$. Inset: we plot the asymmetry $\Delta(f)=\abs{\overline{\tilde{S}_A}(f)-\overline{\tilde{S}_A}(1-f)}$ against the subsystem ratio $f$. Away from $f=1/2$ (finite-size effects), the asymmetry is $\log(\phi)$ as expected.}
    \label{fig:OBC_AEE_J1}
\end{figure}

The asymmetry of the AEE when the total charge is non-abelian is shown in Figure~\ref{fig:OBC_AEE_J1}, where the golden chain Hamiltonian is diagonalized in the $J=1$, $\parity=+1$ sector with $L=24$. Again, we choose to average over 2000 midspectrum eigenvectors. The inset plots the asymmetry $\Delta(f)=\abs{\overline{\tilde{S}_A}(f)-\overline{\tilde{S}_A}(1-f)}$, and shows that the jump discontinuity across $f\leftrightarrow 1-f$ is equal to the predicted value of $\log(\phi)$. At this finite size, the effective crossing of the linear envelopes occurs at $f_*=1/2+1/(2L)\approx 0.5208$.

Overall, we find great agreement of our analytical predictions for Haar-random states and the numerical results for eigenstates of the quantum-chaotic Hamiltonian~\eqref{eq:golden_chain_OBC}\@.

\section{Summary and discussion}
\label{sec:conclusion}
In this work, we developed a theory of typical bipartite entanglement for anyonic systems. Working in a sector of fixed total charge $J$, we computed the statistics for the AEE for Haar-random states in anyonic Hilbert space. The framework provided a quantum group generalization of non-abelian Lie group symmetry-resolved entanglement entropy. Much of our focus was on $\SU(2)_k$ systems, but the assumptions are valid for any UPMC, which covers a vast range of anyon models, including the ones most studied in literature.

We analytically computed the Page curve for Haar-random anyonic states in a fixed topological charge sector. The leading volume-law coefficient is entirely determined by the density of states (\ie the dominant growth rate of the asymptotic fusion space dimension) as expected. In all previous works on Page curves in symmetry sectors, universal corrections like $(f+\log(1-f))/2$ were found even in the $J=0$ case. Symmetry-type terms like these were not found in our case due to the highly constrained Hilbert space structure, in particular, that there are only finitely many spins. For the same reason, the total charge $J$ could not grow extensively with the system size, which meant that the volume-law coefficient was $J$-independent.
We do find a topological-sector term that induces an asymmetry between complementary bipartitions when the total charge is non-abelian, a phenomenon also noted in Ref.~\cite{SREESU(2)} for Lie group symmetries.
To summarize, this shows that constraints coming from fusion categories do not act like Lie group symmetries in the statistics of entanglement.
Despite the presence of superselection sectors and charge conservation, Haar-typical states in anyonic fusion-constrained Hilbert spaces are entanglement-maximal at leading order, in the same universal sense as unconstrained tensor-product systems.

To connect our results back to physical Hamiltonians, we numerically studied the entanglement entropy of highly-excited eigenstates of the integrable and quantum-chaotic golden chain Hamiltonian. We found that the quantum-chaotic model realized the predicted entanglement statistics to leading order, while integrable dynamics deviated from maximality as soon as the subsystem size was extensive. This establishes the anyonic Page curve as the appropriate eigenstate thermalization hypothesis benchmark in topological many-body systems. That is, the AEE acts as a diagnostic for quantum chaotic versus integrable systems. Our results show that Page-type universality extends to a broader class of quantum symmetries than previously understood.

Our work opens up a number of interesting avenues. To our knowledge, this is the first study of typical entanglement entropy beyond the context of Lie group symmetries. This opens the door to studying typical entanglement in the presence of categorical/non-invertible symmetries and in higher-dimensional topological phases. As a first step in this direction, we briefly analyzed the case of anyons on a torus, where the relevant symmetry is given by topological symmetry operators associated with charges threading the torus. This simple example may serve as a starting point for more systematic investigations of typical entanglement in systems with categorical symmetries.

More broadly, we expect the average entanglement we computed to serve as a useful benchmark for generic dynamics in anyonic systems, analogous to the role played by corresponding results in ordinary quantum mechanics. Since relatively little is known about out-of-equilibrium anyonic dynamics, such benchmarks may help identify new phenomena absent in conventional quantum systems.

Finally, our results may also bear on the question of thermalization in systems with symmetries beyond the abelian case. For non-abelian symmetries, seemingly contradictory results have been reported regarding whether the associated symmetry charges hinder~\cite{nyh2016microcanonical,manzano2022non,marvian2022restrictions} or instead enhance thermalization~\cite{potter2016symmetry,majidy2024noncommuting}\@. Investigating whether similar mechanisms arise in the presence of generalized symmetries, including those considered here, would be an interesting direction for future work.

\begin{acknowledgments}
Y.Y.\@ and A.H.\@ thank Daniel Bermudez, Anushya Chandran, J.\ Ignacio Cirac, Nico Kirchner, Frank Pollmann and Cheng-Qian Xu for helpful discussions. L.H. thanks Shayan Majidy and Mario Kieburg for helpful discussions. A.H.\@ thanks Lauritz van Luijk for a valuable regular exchange.

Y.Y.\@ acknowledges support from the
International Max Planck Research School for Quantum Science and Technology (IMPRS-QST). L.H.\@ acknowledges support through an Australian Research Council Australian Discovery Early Career Researcher Award (DECRA) DE230100829 funded by the Australian Government. A.H.\@ acknowledges funding from the BMW Endowment Fund. This research is part of the Munich Quantum Valley (MQV), which is supported by the Bavarian state government with funds from the Hightech Agenda Bayern Plus. This work was made possible through the support of ‘The Quantum Information Structure of Spacetime’ Project (\href{https://www.qiss.fr/}{QISS}) via grant $\#$62312 from the John Templeton Foundation and the `WithOut SpaceTime’ project (\href{https://withoutspacetime.org}{WOST}) via grant $\#$63683 from the John Templeton Foundation. The opinions expressed in this work are those of the author(s) and do not necessarily reflect the views of the John Templeton Foundation.

\end{acknowledgments}

\appendix

\section{Basics of anyon models}
\label{sec:anyon_basics}

In this appendix, we briefly introduce the mathematical basics of anyon models. We start with a very short exposition of the category-theoretical axioms that we use in this paper. Afterwards, we introduce anyons via their graphical calculus of string diagrams. This is followed by a discussion on quantum information aspects of anyon theories.

\subsection{Category-theoretical assumptions}\label{app:category_theory}

For completeness, we summarize the category-theoretical assumptions made in this paper.
For an introduction to category theory from the perspective of anyons, we refer the reader to the following references~\cite{Kitaev_anyon, Bonderson_thesis, anyon_travelogue, Wolf_thesis}\@. For a mathematical introduction, we mention Ref.~\cite{EGNO_book}, in particular chapters~4,~8 and~9. We start with the definition of a fusion category.
\begin{definition}
    A fusion category is a rigid, monoidal, semisimple, $\mathbb{C}$-linear category, such that (i) it has finitely many isomorphism classes of simple objects, (ii) it has finite-dimensional morphism spaces, and (iii) the unit object (with respect to the monoidal structure) is simple.
\end{definition}
For the definition of a rigid monoidal category, see Ref.~\cite[Chapter~2.10]{EGNO_book}\@. For the definition of simple objects and semisimplicity, we refer to~\cite[Def.~1.5.1]{EGNO_book}, and for $\mathbb{C}$-linearity, we refer to Ref.~\cite[Def.~1.2.2]{EGNO_book}\@.

The associators in a fusion category are called $F$-symbols. The $F$-symbols of a fusion category depend on a choice of phase associated with a fusion vertex and are therefore defined only up to gauge transformations~\cite[Section~9.4]{Topological_book}\@. A fusion category is called unitary if there exists a choice of gauge, in which the
F-matrices are unitary. From fusion categories, we can define pre-modular categories.
\begin{definition}
    A pre-modular category (PMC) is a braided fusion category, which has a spherical structure.
\end{definition}
Braided categories are discussed in detail in Ref.~\cite[Chapter~8]{EGNO_book}\@. For the definition of a spherical structure, see Ref.~\cite[Chapter~4.7]{EGNO_book}\@.

Pre-modular categories allow the definition of the $S$-matrix, which we will introduce later in Eq.~\eqref{eq:S-matrix}\@. With this, we are ready to introduce modular tensor categories. See Refs.~\cite{anyon_travelogue,Wolf_thesis} for an extensive review.
\begin{definition}
    A modular tensor category (MTC) is a pre-modular category such that the $S$-matrix is invertible.
\end{definition}
In the following, we will abbreviate a unitary MTC with UMTC and a unitary PMC with UPMC.

The anyons we consider in this paper are described by the simple objects of a UPMC. The set $\mathcal{S}$ of anyons satisfy an abelian, associative fusion algebra $a\otimes b=\bigoplus_c N^c_{ab}c$. Here, $N^c_{ab}\in\mathbb{N}$ are the fusion multiplicities, indicating the number of different ways $a$ and $b$ can fuse to $c$. If $N^c_{ab}\leq 1$ for all anyons $a,b,c\in \mathcal{S}$, then the model is said to be multiplicity-free. There exists a unique vacuum charge $0\in\mathcal{S}$ (unit object) such that $0\otimes a =a\otimes 0 = a$ for any $a\in\mathcal{S}$. For any $a\in\mathcal{S}$, there exists a unique dual charge $\overline{a}\in\mathcal{S}$ such that $N^0_{a\overline{a}}=1$. An anyon $a\in \mathcal{S}$ is said to be non-abelian if $\sum_c N^c_{ab}>1$ for any $b\in\mathcal{S}$, and abelian otherwise. The $\abs{\mathcal{S}}\times\abs{\mathcal{S}}$ fusion matrix 
\begin{equation}
    (N_a)_{bc}=N^{c}_{ab}>0\label{eq:fusion_matrix}
\end{equation}
has a leading positive eigenvalue $d_a$ which is called the quantum dimension of $a$. The quantum dimension also has a graphical definition; see Eq.~\eqref{eq:quantum_loop}\@.

\subsection{Anyonic Hilbert space and graphical calculus}
\label{app:anyon_diagrams}
To each fusion of anyons $a\otimes b\rightarrow c$, there is an associated vector space $V_{ab}^c$ with dimension $\dim V_{ab}^c=N_{ab}^c$ called the fusion space. The state vectors of an anyonic theory are given by fusion vectors, \ie by elements of $V_{ab}^c$.

Anyons possess an extra topological property called isotopy-invariance, meaning that anyon worldlines are invariant under continuous deformations that do not change the topology. While this is obvious from their formulation as a TQFT, it is not apparent from the state-ket notation, so we introduce a diagrammatic notation where isotopy-invariance is natural:
\begin{equation}
    \ket{a\times b\to c}=\left(\frac{d_c}{d_ad_b}\right)^{\frac{1}{4}}\begin{tikzpicture}[baseline={([yshift=-0.5ex]current bounding box.center)},x=\cell cm,y=\cell cm]
        \draw (0,0) node [above left] {\scriptsize $a$} -- (1,-1) -- (1, -2) node [below] {\scriptsize $c$};
        \draw (2,0) node [above right] {\scriptsize $b$} -- (1,-1);
    \end{tikzpicture}\,,
\end{equation}
where the quantum dimension $d_a$ is diagrammatically a loop
\begin{equation}
\label{eq:quantum_loop}
    d_a = \begin{tikzpicture}[baseline={([yshift=-0.5ex]current bounding box.center)},,x=\cell cm,y=\cell cm]
         \draw (0,0) circle (1);
         \node at (1,1) {\scriptsize $a$};
    \end{tikzpicture}\,.
\end{equation}
The normalization factors are chosen so that the categorical $F$-matrices do not need to be altered to represent a change of basis in the diagrammatic notation.

It is common in the physics literature to dress the anyon diagrams with arrows, indicating the worldline of an anyon $a$ (upwards arrow) or the dual anyon $\overline{a}$ (downwards arrow).
Instead, we omit arrows in our notation and directly label each line with the respective object or its dual.

The dual space $V^{ab}_c=(V^c_{ab})^*$ is called splitting space and is spanned by
\begin{equation}
    \bra{a\times b\to c}=\left(\frac{d_c}{d_ad_b}\right)^{\frac{1}{4}}\begin{tikzpicture}[baseline={([yshift=-0.5ex]current bounding box.center)},x=\cell cm,y=\cell cm]
        \draw (0,0) node [below left] {\scriptsize $a$}-- (1,1) -- (1, 2) node [above] {\scriptsize $c$};
        \draw (2,0) node [below right] {\scriptsize $b$} -- (1,1);
    \end{tikzpicture}\,.
\end{equation}
The fusion space of three anyons $a,b,c\in\mathcal{S}$ is constructed as $V^d_{abc}=\bigoplus_x V^x_{ab}\otimes V^d_{xc}$ with basis
\begin{equation}
    \ket{a\times b\to x}\otimes\ket{x\times c\to d}=\left(\frac{d_d}{d_ad_bd_c}\right)^{\frac{1}{4}}
    \begin{tikzpicture}[baseline={([yshift=-0.5ex]current bounding box.center)},x=\cell cm,y=\cell cm]
        \draw (0,0) node [above left] {\scriptsize $a$}-- (2,-2) -- (2,-3) node [below] {\scriptsize $d$};
        \draw (2,0) node [above right] {\scriptsize $b$} -- (1,-1);
        \draw (4,0) node [above right] {\scriptsize $c$} -- (2,-2);
        \node[below left] at (1.7,-1.5) {\scriptsize $x$};
    \end{tikzpicture}\,,
\end{equation}
with $x$ such that $N^x_{ab}\geq 1$, $N_{xc}^d\geq 1$. We could have instead fused $b$ with $c$ first; this gives a fusion space $\bigoplus_y V^y_{bc}\otimes V^d_{ay}$
\begin{equation}
    \ket{b\times c\to y}\otimes \ket{a\times y\to d}=\left(\frac{d_d}{d_ad_bd_c}\right)^{\frac{1}{4}}
    \begin{tikzpicture}[baseline={([yshift=-0.5ex]current bounding box.center)},x=\cell cm,y=\cell cm]
        \draw (0,0) node [above left] {\scriptsize $a$}-- (2,-2) -- (2,-3) node [below] {\scriptsize $d$};
        \draw (2,0) node [above left] {\scriptsize $b$} -- (3,-1);
        \draw (4,0) node [above right] {\scriptsize $c$} -- (2,-2);
        \node[below left] at (3.3,-1.5) {\scriptsize $y$};
    \end{tikzpicture}\,.
\end{equation}
By associativity, the fusion spaces are isomorphic, and the unitary change of basis is given by the so-called $F$-symbol, 
\begin{equation}
    \begin{tikzpicture}[baseline={([yshift=-0.5ex]current bounding box.center)},x=\cell cm,y=\cell cm]
        \draw (0,0) node [above left] {\scriptsize $a$}-- (2,-2) -- (2,-3) node [below] {\scriptsize $d$};
        \draw (2,0) node [above right] {\scriptsize $b$} -- (1,-1);
        \draw (4,0) node [above right] {\scriptsize $c$} -- (2,-2);
        \node[below left] at (1.7,-1.5) {\scriptsize $x$};
    \end{tikzpicture}=\sum_y(F^{abc}_d)_{xy}\begin{tikzpicture}[baseline={([yshift=-0.5ex]current bounding box.center)},x=\cell cm,y=\cell cm]
        \draw (0,0) node [above left] {\scriptsize $a$}-- (2,-2) -- (2,-3) node [below] {\scriptsize $d$};
        \draw (2,0) node [above left] {\scriptsize $b$} -- (3,-1);
        \draw (4,0) node [above right] {\scriptsize $c$} -- (2,-2);
        \node[below left] at (3.3,-1.5) {\scriptsize $y$};
    \end{tikzpicture}\,.
\end{equation}

The fusion space of $L$ anyons $V^J_{a_1\dots a_L}=\bigoplus_{\vb{x}} V^{x_2}_{a_1a_2}\otimes V^{x_3}_{x_2a_3}\otimes\cdots\otimes V^{J}_{x_{L-1}a_L}$ has a standard basis
\begin{equation}
\label{eq:standard_basis_OBC}
    \ket{\vb{x};J}=\left(\frac{d_J}{\prod_{i=1}^L d_{a_i}}\right)^\frac{1}{4}\begin{tikzpicture}[baseline={([yshift=-0.5ex]current bounding box.center)},x=\cell cm,y=\cell cm]
        \draw (0,0) node [above left] {\scriptsize $a_1$} -- (1.6,-1.6);
        \draw (2,0) node [above right] {\scriptsize $a_2$} -- (1,-1);
        \node [above right] at (4,0) {\scriptsize $\cdots$};
        \draw [thick, loosely dotted] (1.6,-1.6) -- (2.4,-2.4);
        \draw (6,0) node [above right] {\scriptsize $a_{L-1}$} -- (3,-3);
        \draw (8,0) node [above right] {\scriptsize $a_L$} -- (4,-4);
        \draw (2.4,-2.4) -- (4,-4) -- (4,-5) node [below] {\scriptsize $J$};
        \node[below left] at (1.7,-1.5) {\scriptsize $x_2$};
        \node[below left] at (3.7,-3.5) {\scriptsize $x_{L-1}$};
    \end{tikzpicture}\,.
\end{equation}
The dimension of fusion space $V^J_{a_1\dots,a_L}$ at fixed charge $J$ is given by $\dim V^J_{a_1\dots,a_L}=\sum_{x_2,x_3\dots,x_{L-1}}N^{x_2}_{a_1a_2}N^{x_3}_{x_2a_3}\cdots N^J_{x_{L-1}a_L}$. Now suppose that $a_1=\cdots=a_L=\mathfrak{j}$, so that one has $\dim V^J_{\mathfrak{j}^{\times L}}=(N^L_\mathfrak{j})_{\mathfrak{j}J}$, where $N_\mathfrak{j}$ is the fusion matrix; see Eq.~\eqref{eq:fusion_matrix}\@. The dimension of the fusion space scales asymptotically like the quantum dimension $\sim d_\mathfrak{j}^L$, with a prefactor depending on $J$. This is the closest analogue of a local Hilbert space dimension, but indeed, $d_\mathfrak{j}$ is irrational in general.

Operators correspond to diagrams in $V^{a_1'\cdots a_L'}_{a_1\cdots a_L}=\bigoplus_J V^J_{a_1\cdots a_L}\otimes V^{a_1'\cdots a_L'}_J$, spanned by
\begin{equation}
    \ketbra{\vb{x};J}{\vb{x}';J}\propto
    \begin{tikzpicture}[baseline={([yshift=-0.5ex]current bounding box.center)},x=\cell cm,y=\cell cm]
        \draw (0,0) node [above left] {\scriptsize $a_1$} -- (1.6,-1.6);
        \draw (2,0) node [above right] {\scriptsize $a_2$} -- (1,-1);
        \node [above right] at (4,0) {\scriptsize $\cdots$};
        \draw [thick, loosely dotted] (1.6,-1.6) -- (2.4,-2.4);
        \draw (6,0) node [above right] {\scriptsize $a_{L-1}$} -- (3,-3);
        \draw (8,0) node [above right] {\scriptsize $a_L$} -- (4,-4);
        \draw (2.4,-2.4) -- (4,-4) -- (4,-6);
        \node [right] at (4,-5) {\scriptsize $J$};
        \node[below left] at (1.7,-1.5) {\scriptsize $x_2$};
        \node[below left] at (3.7,-3.5) {\scriptsize $x_{L-1}$};
        \draw (0,-10) node [below left] {\scriptsize $a_1'$}-- (1.6,-8.4);
        \draw (2,-10) node [below right] {\scriptsize $a_2'$} -- (1,-9);
        \node [below right] at (4,-10) {\scriptsize $\cdots$};
        \draw [thick, loosely dotted] (1.6,-8.4) -- (2.4,-7.6);
        \draw (6,-10) node [below right] {\scriptsize $a_{L-1}'$} -- (3,-7);
        \draw (8,-10) node [below right] {\scriptsize $a_L'$} -- (4,-6);
        \draw (2.4,-7.6) -- (4,-6);
        \node[above left] at (1.5,-8.5) {\scriptsize $x_2'$};
        \node[above left] at (3.5,-6.5) {\scriptsize $x_{L-1}'$};
    \end{tikzpicture}\,.
\end{equation}
Notice that only charge-conserving operators exist in the anyonic Hilbert space. This is a physical constraint of the theory, called charge superselection (CSS), which we discuss in more detail in Appendix~\ref{app:superselection}\@. Every operator $O\in V^{a_1'\cdots a_L'}_{a_1\cdots a_L}$ can be projected onto a charge-$J$ sector with the projector $\Pi_J$. We will then write $[O]_J=\Pi_J O\Pi_J\in V^J_{a_1\cdots a_L}\otimes V^{a_1'\cdots a_L'}_J$. Vice versa, $O=\bigoplus_{J\in\mathcal{S}} [O]_J$.

For calculations, we will find it much more convenient to use a basis adapted to a bipartition into systems $A$ and $B$, containing $L_A$ anyons $\{a_1,\dots,a_{L_A}\}$ and $L_B=L-L_A$ anyons $\{a_{L_A+1},\dots,a_{L}\}$ respectively:
\begin{equation}
\label{eq:bipartite_basis}
    \ket{\vb{x},\vb{y};J}\propto\begin{tikzpicture}[baseline={([yshift=-0.5ex]current bounding box.center)},x=\cell cm,y=\cell cm]
        \draw (0,0) node [above left] {\scriptsize $a_1$} -- (1.6,-1.6);
        \draw (2,0) node [above right] {\scriptsize $a_2$} -- (1,-1);
        \node [above right] at (4,0) {\scriptsize $\cdots$}; 
        \draw [thick, loosely dotted] (1.6,-1.6) -- (2.4,-2.4);
        \node[below left] at (1.5,-1.2) {\scriptsize $x_2$};
        \node [below] at (1.7, -2.6) {\scriptsize $x_{L_A-1}$};
        \draw (6,0) node [above right] {\scriptsize $a_{L_A}$} -- (3,-3);
        \draw (10,0) -- (13,-3);
        \node [above] at (9.33,0) {\scriptsize $a_{L_A+1}$};
        \node [above] at (11.47,0) {\scriptsize $\cdots$};
        \draw (14,0) -- (15,-1);
        \node [above] at (13.33,0) {\scriptsize $a_{L-1}$};
        \draw (16,0) node [above right] {\scriptsize $a_L$} -- (14.4,-1.6);
        \draw (2.4,-2.4) -- (3,-3) --(8,-4);
        \draw [thick, loosely dotted] (14.4,-1.6) -- (13.6,-2.4);
        \node[below right] at (14.7,-1.2) {\scriptsize $y_2$};
        \node [below] at (14.5, -2.6) {\scriptsize $y_{L_B-1}$};
        \draw (13.6,-2.4) -- (13,-3) -- (8,-4);
        \draw (8,-5) -- (8,-4);
        \node [right] at (8,-4.5) {\scriptsize $J$};
        \node at (5.6,-4.2) {\scriptsize $x_{L_A}$};
        \node at (10.8,-4.2) {\scriptsize $y_{L_B}$};
    \end{tikzpicture}
\end{equation}
In this way, we can talk about the charge in $A$, $\alpha=x_{L_A}$, and the charge in $B$, $\beta=y_{L_B}$, satisfying $N^J_{\alpha\beta}=1$. The fusion space adapted to this basis decomposes as $\bigoplus_{\alpha,\beta} V_{a_1\dots a_{L_A}}^\alpha\otimes V_{a_{L_A+1}\dots a_L}^\beta \otimes V^J_{\alpha\beta}$. The bases \eqref{eq:standard_basis_OBC} and \eqref{eq:bipartite_basis} are related by several $F$-moves. Since they are unitary basis changes, they do not affect the entanglement entropy.

Another operation available to anyonic systems is the braiding of two neighboring anyons, producing non-trivial exchange statistics. Diagrammatically, the braiding of two anyons $a,b$ is a two-local operator $B$ acting on basis vectors as
\begin{equation}
    \begin{tikzpicture}[baseline={([yshift=-0.5ex]current bounding box.center)},x=\cell cm,y=\cell cm]]
        \draw (-1.5,0) -- (3,0);
        \begin{knot}[clip width=3]
            \strand (0,0) .. controls +(0,1) and +(0,-1) .. (1.5,2) node [above] {\scriptsize $a$};
            \strand (1.5,0) .. controls +(0,1) and +(0,-1) .. (0,2) node [above] {\scriptsize $b$};;
        \end{knot}
        \node[below] at (-0.75,0) {\scriptsize $x_{i-1}$};
        \node[below] at (0.75,0) {\scriptsize $x_i$};
        \node[below] at (2.25,0) {\scriptsize $x_{i+1}$};
    \end{tikzpicture}\hspace{-4mm}
    =
    \sum_{x_i'} \big(B_{x_{i+1}}^{x_{i-1}ba}\big)_{x_ix_i'}
    \begin{tikzpicture}[baseline={([yshift=-0.5ex]current bounding box.center)},x=\cell cm,y=\cell cm]]
        \draw (-1.5,0) -- (3,0);
        \draw (0,0) -- (0,2) node [above] {\scriptsize $b$};
        \draw (1.5,0) -- (1.5,2) node [above] {\scriptsize $a$};
        \node[below] at (-0.75,0) {\scriptsize $x_{i-1}$};
        \node[below] at (0.75,0.2) {\scriptsize $x_i'$};
        \node[below] at (2.25,0) {\scriptsize $x_{i+1}$};
    \end{tikzpicture}\,.
\end{equation}
Its inverse is given by reversing the crossing,
\begin{equation}
    \begin{tikzpicture}[baseline={([yshift=-0.5ex]current bounding box.center)},x=\cell cm,y=\cell cm]]
        \draw (-1.5,0) -- (3,0);
        \begin{knot}[clip width=3, flip crossing=1]
            \strand (0,0) .. controls +(0,1) and +(0,-1) .. (1.5,2) node [above] {\scriptsize $a$};
            \strand (1.5,0) .. controls +(0,1) and +(0,-1) .. (0,2) node [above] {\scriptsize $b$};;
        \end{knot}
        \node[below] at (-0.75,0) {\scriptsize $x_{i-1}$};
        \node[below] at (0.75,0) {\scriptsize $x_i$};
        \node[below] at (2.25,0) {\scriptsize $x_{i+1}$};
    \end{tikzpicture}\hspace{-4mm}
    =
    \sum_{x_i'} \big(B_{x_{i+1}}^{x_{i-1}ba}\big)^{-1}_{x_ix_i'}
    \begin{tikzpicture}[baseline={([yshift=-0.5ex]current bounding box.center)},x=\cell cm,y=\cell cm]]
        \draw (-1.5,0) -- (3,0);
        \draw (0,0) -- (0,2) node [above] {\scriptsize $b$};
        \draw (1.5,0) -- (1.5,2) node [above] {\scriptsize $a$};
        \node[below] at (-0.75,0) {\scriptsize $x_{i-1}$};
        \node[below] at (0.75,0.2) {\scriptsize $x_i'$};
        \node[below] at (2.25,0) {\scriptsize $x_{i+1}$};
    \end{tikzpicture}\,,
\end{equation}
and algebraically relates to $B$ via
\begin{equation}
    \big(B_d^{abc}\big)^{-1}_{ef} = \big(\overline{B_d^{abc}}\big)_{fe}.
\end{equation}
The braiding is related to the $R$-symbol, which on fusion vertices has the action
\begin{equation}
    \begin{tikzpicture}[baseline={([yshift=-0.5ex]current bounding box.center)},x=\cell cm,y=\cell cm]
    \draw (-1,1) -- (0,0) -- (0,-1) node [below] {\scriptsize $c$};
    \draw (1,1) -- (0,0);
    \begin{knot}[clip width=3]
    \strand (-1,1) .. controls +(0,1) and +(0,-1) .. (1,3) node [above] {\scriptsize $a$};
    \strand (1,1) .. controls +(0,1) and +(0,-1) .. (-1,3) node [above] {\scriptsize $b$};
    \end{knot}
    \end{tikzpicture}=R^{ab}_c\begin{tikzpicture}[baseline={([yshift=-0.5ex]current bounding box.center)},x=\cell cm,y=\cell cm]
    \draw (-1,1) node [above left] {\scriptsize $b$} -- (0,0) -- (0,-1) node [below] {\scriptsize $c$};
    \draw (1,1) node [above right] {\scriptsize $a$} -- (0,0);
    \end{tikzpicture}\,,
\end{equation}
where, in the case of multiplicity-free UPMCs, $R^{ab}_c\in \mathrm{U}(1)$ is a complex phase. In particular, we have the relation
\begin{equation}
    \big(B_d^{abc}\big)_{ef} = \sum_g \big(F_d^{acb}\big)_{eg} R_g^{cb} \big(F_d^{abc}\big)^{-1}_{gf}.
\end{equation}

Finally, the braiding allows us to introduce a crucial object, called the topological / modular $S$-matrix. It is part of the modular data of the UMTC and encodes the quantum dimensions and fusion space dimensions of the theory. Its matrix elements are given by the Hopf link
\begin{equation}
\label{eq:S-matrix}
    S_{ab}=\frac{1}{\mathcal{D}}\ \begin{tikzpicture}[baseline={([yshift=-0.5ex]current bounding box.center)},x=\cell cm,y=\cell cm]
    \begin{knot}[clip width=3, flip crossing=1]
    \strand (0,0) circle (1);
    \strand (1,0) circle (1);
    \end{knot}
    \node at (-1,1) {\scriptsize $a$};
    \node at (2,1) {\scriptsize $b$};
    \end{tikzpicture}=\frac{1}{\mathcal{D}}\sum_c d_c R^{b\overline{a}}_cR^{\overline{a}b}_b\,,
\end{equation}
where $\mathcal{D}=\sqrt{\sum_a d_a^2}$ is the total quantum dimension. In a UMTC, the $S$-matrix is unitary and symmetric, and can be interpreted as a change of basis for the fusion tree on a torus \cite{Pfeifer_PBC}\@. For us, it will be enough to quote the famous Verlinde formula, relating the fusion coefficients to the $S$-matrix:
\begin{equation}
    N^c_{ab}=\sum_x \frac{S_{ax}S_{bx}S_{cx}^*}{S_{1x}}\,.
\end{equation}
It is used to calculate the scaling of the Hilbert space dimensions for modular theories in Section~\ref{sec:large-L} and Appendix~\ref{app:PBC}\@.

\subsection{The quantum trace}\label{app:quantum_trace}

Next, we introduce the trace on the anyonic Hilbert space. Notice that for theories with observable algebras that are direct sums of matrix algebras, we have the freedom to rescale the trace. This can be done by introducing a rescaling factor in each superselection sector. See, for instance, Ref.~\cite{RT-QEC}\@. The rescaling is not uniquely defined from the algebraic perspective. In our case, however, we have to ensure that the trace is consistent with planar isotopy invariance. This fixes the trace uniquely to the quantum trace~\cite{Topological_book}\@. We follow Ref.~\cite{Bonderson_entanglement} and denote the quantum trace by $\tilde{\tr}$. The quantum trace of an operator $O\in V_{a'_1\dots a'_L}^{a_1\dots a_L}$ is related to the normal Hilbert space trace $\tr$ by
\begin{align}
    \tilde{\tr}(O) &= \sum_J d_J \tr([O]_J),\label{eq:quantum_trace_1}\\
    \tr(O) &= \sum_J \frac{1}{d_J} \tilde{\tr}([O]_J),\label{eq:quantum_trace_2}\,,
\end{align}
\ie a factor of the quantum dimension within each sector. Diagrammatically, the quantum trace corresponds to closing all external legs of a diagram, for instance,
\begin{equation}
    \tilde{\tr}\left(\begin{tikzpicture}[baseline={([yshift=-0.5ex]current bounding box.center)},x=\cell cm,y=\cell cm]
        \draw (0,0) node [above left] {\scriptsize $a$} -- (1,-1) --node [right] {\scriptsize $c$}  (1, -2) ;
        \draw (2,0) node [above right] {\scriptsize $b$} -- (1,-1);
        \draw (1,-2) -- (0,-3) node[below left] {\scriptsize $a$};
        \draw (1,-2) -- (2,-3) node[below right] {\scriptsize $b$};
    \end{tikzpicture}\right)
    =
    \begin{tikzpicture}[baseline={([yshift=-0.5ex]current bounding box.center)},x=\cell cm,y=\cell cm]
        \draw (0,0) -- (1,-1) --node [right] {\scriptsize $c$}  (1, -2) ;
        \draw (2,0) -- (1,-1);
        \draw (1,-2) -- (0,-3) -- (-1,-2) -- node[left] {\scriptsize $a$} (-1,-1) -- (0,0);
        \draw (1,-2) -- (2,-3) -- (3,-2) -- node[right] {\scriptsize $b$} (3,-1) -- (2,0);
    \end{tikzpicture}\,.
\end{equation}

To define a partial trace on anyonic Hilbert space, we first choose a basis $\ket{\vb{x},\alpha,\vb{y};J}$ adapted to the bipartition as in Eq.~\eqref{eq:bipartite_basis}. The standard partial trace over $B$
\begin{equation}
    \tr_B \ketbra{\vb{x},\alpha,\vb{y};J}{\vb{x}',\alpha',\vb{y}';J}=\delta_{\alpha,\alpha'}\delta_{\vb{y},\vb{y}'}\ketbra{\vb{x};\alpha}{\vb{x}';\alpha}
\end{equation}
is not invariant under planar isotopy. For example, on tensor products, $\tr_B(O_A\otimes O_B)=\sum_{a,b,c} N^c_{ab}\tr([O_B]_b)[O_A]_a$. As before, to retain isotopy-invariance, we use the categorical version of the partial trace $\tilde{\tr}_B$, defined by closing the external legs of the diagram pertaining to system $B$ and using diagrammatic calculus to evaluate it. Equivalently, it is related to the standard trace via
\begin{equation}
    \tilde{\tr}_B\,O=\sum_b \frac{d_b}{d_c} \big[\tr_B[O]_b\big]_c\,.
\end{equation}
This partial quantum trace is the one that behaves as expected under tensor products, $\tilde{\tr}_B(O_A\otimes O_B)=\tilde{\tr}(O_B) O_A$.

\subsection{Superselection and operator algebras}
\label{app:superselection}
Out of all operators on Hilbert space, anyonic theories only admit a subset of charge-conserving operators. This phenomena is called charge superselection (CSS). It is a physical or operational constraint, imposed by the microscopic physics underlying the emergent anyonic particles. Here we will be agnostic to the microscopic physics, but explore the implications of a such a constraint on the anyonic theory. 

CSS is most naturally described in the algebraic formulation of quantum mechanics, where physical observables form a von Neumann algebra. In the finite-dimensional setting, the theory reduces to the study of matrix algebras.

Suppose $\mathcal{H}$ is a finite-dimensional Hilbert space and denote the algebra of bounded operators on $\mathcal{H}$ by $\mathcal{B}(\mathcal{H})$. Let $\mathcal{M}\subseteq \mathcal{B}(\mathcal{H})$ be a von Neumann algebra acting on $\mathcal{H}$. We interpret $\mathcal{M}$ as the algebra of observables accessible to an agent in the system. For a standard quantum many-body system living in $\mathcal{H}$, one usually takes the full matrix algebra $\mathcal{M}=\mathcal{B}(\mathcal{H})$.

On the other hand, systems exhibiting CSS have a restricted algebra of observables. One such restriction, which was studied in Ref.~\cite{SREESU(2)}, is to impose that all allowed operations commute with a symmetry group action $G$, the so-called $G$-invariant observables. Under a unitary representation of $G$, the Hilbert space $\mathcal{H}$ decomposes as $\mathcal{H}\cong \bigoplus_\lambda V_\lambda\otimes \mathbb{C}^{m_\lambda}$, where $V_\lambda$ is the irreducible representation labeled by $\lambda$ and $m_\lambda$ is its multiplicity in the decomposition. The $G$-invariant observable algebra is given by $\mathcal{M}=\bigoplus_\lambda I_{V_\lambda}\otimes \mathcal{B}(\mathbb{C}^{m_\lambda})$, where $I_{V_\lambda}$ is the identity operator on $V_\lambda$. 

For anyons, the Hilbert space naturally decomposes into total charge sectors $\mathcal{H}=\bigoplus_J V^J_{a_1\dots a_L}$ and one takes $\mathcal{M}=\bigoplus_J \mathcal{B}(V^J_{a_1\dots a_L})$ to be the full matrix algebra on each $J$ sector. Physically, that means that every available operation in $\mathcal{M}$ has to respect the direct sum structure in $J$ and cannot mix between different blocks.

An immediate consequence of CSS is that superpositions of pure states in different charge sectors are operationally indistinguishable from mixtures of these states. That is, if $\ket{J}\in V^J_{a_1\dots a_L}$ and $\ket{K}\in V^K_{a_1\dots a_L}$, then $\alpha \ket{J}+\beta\ket{K}$ produces the same expectation values as $\tilde{\rho}=\abs{\alpha}^2 /d_J\ketbra{J}{J}+\abs{\beta}^2/d_K\ketbra{K}{K}$ with respect to any $A\in\mathcal{M}$. One often says that $\ket{J}$ and $\ket{K}$ are not superposable or incoherent.

\subsection{Anyonic states}
\label{app:pure_states}
Anyonic density operators $\tilde{\rho}\in \mathcal{M}$ are semi-positive definite operators on anyonic Hilbert space normalized with respect to the quantum trace, $\tilde{\tr}\tilde{\rho}=1$. To achieve this normalization, we divide by the quantum dimension of the total charge $d_J$ in each sector, \ie
\begin{equation}
    \tilde{\rho} = \sum_J \frac{1}{d_J} \sum_{\vb{x},\vb{x}'} c_{\vb{x}'}c_{\vb{x}}^*\ketbra{\vb{x};J}{\vb{x}';J}.
\end{equation}

As mentioned in the main text, there is some current discussion on whether a state $\tilde{\rho}=1/d_J\ketbra{J}{J}$ with non-abelian total charge $J$ can be pure. On the one hand, such a state is represented as a rank-1 projector, making it extremal in the convex set of states. On the other hand, such a state admits properties of a mixed state, such as non-vanishing AEE or anyonic purity $\tilde{\tr}(\tilde{\rho}^2)$ smaller than one. Since the property of a state being pure has an operational meaning, this debate can be tackled by taking an operational perspective. 

We define anyonic states to be pure if they are extremal with respect to a specified algebra of observables $\mathcal{M}$. For the anyonic Hilbert space, we will always take $\mathcal{M}=\bigoplus_J\mathcal{B}(\mathcal{H}^J)$---all charge-conserving operators. Then it is simple to show that a state $\tilde{\rho}\in\mathscr{S}(\mathcal{M})$ is pure if and only if it is supported in a single $J$ sector and is rank-1 (see Lemma~\ref{lem:pure_states}). However, even among these pure states, we find that those with non-abelian total charge $J$ contain information content in the limit of many copies. This means that a faithful measure of entanglement eludes these states.

While we do not provide a resolution to the problem, nor do we attempt to, we elucidate the structure of anyonic states by identifying three classes: i) pure, with abelian total charge $J$; ii) pure, with non-abelian total charge $J$; iii) mixed.
These have also been recognized in~\cite{Kato_entanglement}, see Appendix~B therein. We now show from first principles how these classes arise.

Consider the Hilbert space $\mathcal{H}^0=\mathrm{span}_{\vb{x},J}\{\ket{\vb{x},J;0}\}$ spanned by vectors $\ket{\vb{x},J;0}$, where $L+1$ anyons fuse to the trivial charge. The fusion constraint of trivial total charge fixes anyon $a_{L+1}$ to be the dual of the last intermediate fusion label, which we single out and denote by $J$. Graphically, these vectors are represented by
\begin{equation}
    \ket{\vb{x},J;0} \propto \begin{tikzpicture}[baseline={([yshift=-0.5ex]current bounding box.center)},x=\cell cm,y=\cell cm]
        \draw (0,0) node [above left] {\scriptsize $a_1$} -- (1.6,-1.6);
        \draw (2,0) node [above right] {\scriptsize $a_2$} -- (1,-1);
        \node [above right] at (4,0) {\scriptsize $\cdots$};
        \draw [thick, loosely dotted] (1.6,-1.6) -- (2.4,-2.4);
        \draw (6,0) node [above right] {\scriptsize $a_{L}$} -- (3,-3);
        \draw (8,0) node [above right] {\scriptsize $\overline{J}$} -- (4,-4);
        \draw (2.4,-2.4) -- (4,-4);
        \draw[dashed] (4,-4) -- (4,-5) node [below] {\scriptsize $0$};
        \node[below left] at (1.7,-1.5) {\scriptsize $\vb{x}$};
        \node[below left] at (3.7,-3.5) {\scriptsize $J$};
    \end{tikzpicture}\,.\label{eq:pure_chain}
\end{equation}
Then, the von Neumann algebra acting on $\mathcal{H}^0$ is given by $\mathcal{M}_0=\mathcal{B}(\mathcal{H}^0)$. Clearly, the states induced by vectors of the form of Eq.~\eqref{eq:pure_chain} are pure states on $\mathcal{M}^0$: They are vector states with an abelian total charge of $0$. Notice that this even holds for states, which are linear combinations over different $J$. Restricting to a fixed $J$ sector for now, we can form a density operator from vectors of the form of Eq.~\eqref{eq:pure_chain}, $[\tilde{\rho}_{0}]_J=1/d_J\sum_{\vb{x},\vb{x}'}\Psi_{\vb{x}}\Psi_{\vb{x}'}^*\ketbra{\vb{x},J;0}{\vb{x}',J;0}$, where the $\Psi_{\vb{x}}\in\mathbb{C}$ are complex coefficients. We can easily verify graphically that tracing over $\overline{J}$ gives a reduced density operator $\tilde{\tr}_{\overline{J}}[\tilde{\rho}_{0}]_J=\tilde{\rho}_J=1/d_J\sum_{\vb{x},\vb{x}'}\Psi_{\vb{x}}\Psi_{\vb{x}'}^*\ketbra{\vb{x};J}{\vb{x'};J}$, where
\begin{equation}
    \ket{\vb{x};J} \propto \begin{tikzpicture}[baseline={([yshift=-0.5ex]current bounding box.center)},x=\cell cm,y=\cell cm]
        \draw (0,0) node [above left] {\scriptsize $a_1$} -- (1.6,-1.6);
        \draw (2,0) node [above right] {\scriptsize $a_2$} -- (1,-1);
        \node [above right] at (4,0) {\scriptsize $\cdots$};
        \draw [thick, loosely dotted] (1.6,-1.6) -- (2.4,-2.4);
        \draw (6,0) node [above right] {\scriptsize $a_{L}$} -- (3,-3);
        \draw (2.4,-2.4) -- (4,-4);
        \node[below left] at (1.7,-1.5) {\scriptsize $\vb{x}$};
        \node[below left] at (3.7,-3.5) {\scriptsize $J$};
    \end{tikzpicture}\,.\label{eq:pure_chain_non_abelian}
\end{equation}
For each $J$, the vectors $\ket{\vb{x};J}$ span a Hilbert space $\mathcal{H}^J=\mathrm{span}_{\vb{x}}\{\ket{\vb{x};J}\}$. Notice that we can obtain the vectors $\ket{\vb{x};J}$ from $\ket{\vb{x},J;0}$ by just bending down the $\overline{J}$ charge using planar isotopy-invariance. Algebraically, this corresponds to the fact that $\bigoplus_J \mathcal{H}^J\otimes V_{J\overline{J}}^0\simeq \bigoplus_J \mathcal{H}^J$ because each $V_{J\overline{J}}^0$ is one-dimensional. We immediately see that $\mathcal{M}_0\simeq\mathcal{B}(\bigoplus_J \mathcal{H}^J\otimes V_{J\overline{J}}^0)\simeq\mathcal{B}(\bigoplus_J \mathcal{H}^J)$. Furthermore, the von Neumann algebra acting on vectors of the form~\eqref{eq:pure_chain_non_abelian} is given by $\mathcal{M}_\mathrm{sub} = \bigoplus_J \mathcal{B}(\mathcal{H}^J)$. Even though $\mathcal{M}_0$ and $\mathcal{M}_\mathrm{sub}$ act on isomorphic Hilbert spaces, they are not isomorphic as $C^*$-algebras. This can easily be seen by considering their respective centres: $\mathcal{Z}(\mathcal{M}_0)=\mathbb{C}\mathds{1}$, while $\mathcal{Z}(\mathcal{M}_\mathrm{sub})=\{\Pi_J\}$, \ie\!\!, the projectors onto the charge-$J$ sector. In fact, we have $\mathcal{M}_\mathrm{sub}\subsetneq\mathcal{M}_0$, and there is an injective $*$-homomorphism $\iota:\mathcal{M}_\mathrm{sub}\hookrightarrow\mathcal{M}_0$ realized by block-diagonal inclusion.
We will now show that the restriction of a pure state in $\mathscr{\mathcal{M}_0}$ to a state in $\mathscr{\mathcal{M}_\mathrm{sub}}$ via $\iota$ remains pure. 
\begin{lemma}\label{lem:pure_states}
    Take a fixed $J$ sector, say $J_0$, and define the vector $\ket{\psi}=\ket{\vb{x},J_0;0}$ in $\mathcal{H}^0$ and corresponding vector state $\omega\in\mathscr{S}(\mathcal{M}_0)$ by $\omega(O)=\langle\psi, O\psi\rangle$, $O\in\mathcal{M}_0$.
    Restrict $\omega$ to $\mathcal{M}_\mathrm{sub}$ by defining $\omega_\mathrm{sub}=\omega\circ\iota$. Then $\omega_\mathrm{sub}\in\mathscr{S}(\mathcal{M}_\mathrm{sub})$ is a pure state with respect to $\mathcal{M}_\mathrm{sub}$.
    \begin{proof}
         We notice that any state $\phi\in\mathscr{S(\mathcal{M}_\mathrm{sub})}$ decomposes as $\phi(\bullet)=\sum_J\lambda_J\phi_J(\bullet)$, where $\lambda_J=\phi(\Pi_J)$ is the probability to be in block $J$ and $\phi_J$ is a state on a single block $\mathcal{B}(\mathcal{H}^J)$. Such a state on a direct sum algebra $\bigoplus_J\mathcal{B}(\mathcal{H}^J)$ is pure if and only if
\begin{enumerate}[(i)]
    \item it has support only on one block $J_0$: $\lambda_{J_0}=1$, and all other $\lambda_J=0$, and
    \item the induced state $\phi_{J_0}$ on $\mathcal{B}(\mathcal{H}^{J_0})$ is pure.
\end{enumerate}
In the case of $\omega_\mathrm{sub}$, we conclude
\begin{equation}
    \omega_\mathrm{sub}(\Pi_J) = \omega\circ\iota(\Pi_J) = \langle \psi,\Pi_J\psi\rangle = \begin{cases}
        1,\quad J=J_0,\\
        0,\quad J\neq J_0.
    \end{cases}
\end{equation}
Hence, (i) is satisfied. Furthermore, when restricted to $\mathcal{B}(\mathcal{H}^{J_0})$, $[\omega_{\mathrm{sub}}]_{J_0}(O)=\langle\psi,O\psi\rangle$, $O\in\mathcal{B}(\mathcal{H}^{J_0})$. Therefore, $[\omega_{\mathrm{sub}}]_{J_0}$ is a vector state on the full matrix algebra $\mathcal{B}(\mathcal{H}^{J_0})$, and hence (ii) is also satisfied.
    \end{proof}
\end{lemma}
    
Even though a state $\tilde{\rho}_J=1/d_J\ketbra{\vb{x};J}{\vb{x}';J}$ with a fixed non-abelian total charge $J$ is pure by Lemma~\ref{lem:pure_states}, it nevertheless has information content. This can be seen by calculating the AEE, $\tilde{S}_{AB}(\tilde{\rho}_J) = \tilde{\tr}(\tilde{\rho}_J\log\tilde{\rho}_J) = \log(d_J)\neq 0$. The reason for this is the presence of CSS, which allows pure states to exhibit entanglement properties that are usually only present for mixed states~\cite{entanglement_superselection_mixedness}\@. In particular, CSS can lead to bound entanglement and activation for pure states. This is precisely what happens for $\tilde{\rho}_J$: while the distillable entanglement for $\tilde{\rho}_J$ using local operations and classical communication (LOCC) from $\mathcal{M}_\mathrm{sub}$ is zero (because it is pure), entanglement distillation becomes possible when when additional copies of $\tilde{\rho}_J$ are available. To see this, consider two copies of the two-anyon density operator $\tilde{\rho}_J=1/d_J\ketbra{J}{J}$ with non-abelian total charge $J$:
\begin{equation}
    \begin{tikzpicture}[baseline={([yshift=-0.5ex]current bounding box.center)},x=\cell cm,y=\cell cm]
    \draw (-1,1) node [above left] {\scriptsize $a_1$} -- (0,0) -- node [right] {\scriptsize $J$} (0,-1);
    \draw (1,1) node [above right] {\scriptsize $a_2$} -- (0,0);
    \draw (0,-1) -- (-1,-2) node[below left] {\scriptsize$a_1$};
    \draw (0,-1) -- (1,-2) node[below right] {\scriptsize $a_2$};
    \end{tikzpicture}\hspace{-2mm}
    \otimes \hspace{-2mm}
    \begin{tikzpicture}[baseline={([yshift=-0.5ex]current bounding box.center)},x=\cell cm,y=\cell cm]
    \draw (-1,1) node [above left] {\scriptsize $a_1$} -- (0,0) -- node [right] {\scriptsize $J$} (0,-1);
    \draw (1,1) node [above right] {\scriptsize $a_2$} -- (0,0);
    \draw (0,-1) -- (-1,-2) node[below left] {\scriptsize$a_1$};
    \draw (0,-1) -- (1,-2) node[below right] {\scriptsize $a_2$};
    \end{tikzpicture}\hspace{-3mm}
    =\bigoplus_{c\mid N^c_{JJ}=1}\frac{\sqrt{d_c}}{d_J}\hspace{-5mm}
    \begin{tikzpicture}[baseline={([yshift=-0.5ex]current bounding box.center)},x=\cell cm,y=\cell cm]
        \draw (0,0) node [above left] {\scriptsize $a_1$} -- (1,-1) -- (3.5,-2) -- (6,-1) -- (7,0) node [above right] {\scriptsize $a_2$};
        \draw (2,0) node [above right] {\scriptsize $a_2$} -- (1,-1);
        \draw (5,0) node [above left] {\scriptsize $a_1$} -- (6,-1);
        \draw (3.5,-2) -- (3.5,-3);
        \node at (2,-2) {\scriptsize $J$};
        \node at (5,-2) {\scriptsize $J$};
        
        \node at (4.1,-2.5) {\scriptsize $c$};

        \begin{scope}[shift={(0,-0.5)}, yscale=-1, shift={(0,4.5)}]
        \draw (0,0) node [below left] {\scriptsize $a_1$} -- (1,-1) -- (3.5,-2) -- (6,-1) -- (7,0) node [below right] {\scriptsize $a_2$};
        \draw (2,0) node [below right] {\scriptsize $a_2$} -- (1,-1);
        \draw (5,0) node [below left] {\scriptsize $a_1$} -- (6,-1);
        \node at (2,-2) {\scriptsize $J$};
        \node at (5,-2) {\scriptsize $J$};
        \end{scope}
    \end{tikzpicture}\,.
\end{equation}
Because $J$ is non-abelian, there are multiple possible fusion labels $c$. Hence, even though $\tilde{\rho}_J$ is pure, $\tilde{\rho}_J\otimes\tilde{\rho}_J$ is mixed. The additional fusion spaces allow relational information between copies of $\tilde{\rho}_J$ to be accessed by local operations respecting CSS. Thus, multiple copies can activate otherwise bound superselection-induced entanglement~\cite{superactivation_anyon}\@.

Finally, for the third class of states---mixed states---coming up with an entanglement measure in anyonic Hilbert space requires, at the very least, an entanglement measure for the mixed states in standard Hilbert spaces. We do not discuss mixed-state bipartite entanglement here.

\subsection{Anyons on higher-genus surfaces}\label{app:PBC}
We make a short comment on the fusion spaces of anyons placed on surfaces with nonzero genus $g$. We focus on the case with $g=1$, \ie a torus, but the structures here can easily be generalised to higher genera. 

First, we prescribe a basis for the toric fusion space. We begin by reminding ourselves where the fusion tree basis for a disk comes from. The $(2+1)D$ TQFT description says that each free anyon leg in a fusion basis is equivalent to a puncture on a 2-sphere: 
\begin{equation}
\label{eq:fusion_tree}
    \begin{tikzpicture}[baseline={([yshift=-0.5ex]current bounding box.center)},x=\cell cm,y=\cell cm]]
        \draw (-1.5,0) -- (3,0);
        \draw (0,0) -- (0,2) node [above] {\scriptsize $a_1$};
        \draw (1.5,0) -- (1.5,2) node [above] {\scriptsize $a_2$};
        \node[below] at (-0.75,0) {\scriptsize $x_0$};
        \node[below] at (0.75,0) {\scriptsize $x_1$};
        \node[below] at (2.25,0) {\scriptsize $x_2$};
        \draw (4.5,0) -- (7.5,0);
        \draw (6,0) -- (6,2) node [above] {\scriptsize $a_L$};
        \node[below] at (5.25,0) {\scriptsize $x_{L-1}$};
        \node[below] at (6.75,0) {\scriptsize $x_L$};
        \node at (3.75,0) {\scriptsize $\cdots$};
    \end{tikzpicture}\quad \cong \ \quad \begin{tikzpicture}[baseline={([yshift=-0.5ex]current bounding box.center)}]
        \draw [fill=black!10] (0,0) circle (1);
        \draw [fill=white] (0,0.8) node [circle,fill=black,inner sep=1pt] {} ellipse (0.25 and 0.1);
        \draw [dashed] (0,-0.8) node [circle,fill=black,inner sep=1pt] {} ellipse (0.25 and 0.1);
        \draw [dashed] (0,0) ellipse (1 and 0.4); 

        \draw (0,0.8) -- (0,0.352);
        \draw [dotted] (0,0.352) -- (0,-0.032);
        \draw (0,-0.032) -- (0,-0.8);
        \draw (-0.64, 0.48) node [circle,fill=black,inner sep=1pt] {} -- (0,0.48);
        \draw (-0.784, -0.16) node [circle,fill=black,inner sep=1pt] {} -- (0,-0.16);
        \draw (-0.64, -0.48) node [circle,fill=black,inner sep=1pt] {} -- (0,-0.48);
    \end{tikzpicture}\,.
\end{equation}
We enlarged the punctures carrying charges $x_0$ and $x_L$ in preparation for the gluing procedure to come. To recover the fusion tree for anyons on a disk, one takes $x_0=0$ and calls $J=x_L$ the topological charge. Thus, we have $L+1$ punctures on the sphere, which is homeomorphic to $L$ punctures on the disk. This recovers the description of $L$ anyons on a disk with a charge on the boundary.

To obtain a basis for the fusion space of anyons on a torus, we ask for the ends $x_0=x_L$ to be identified, and simply call this label $x$. Topologically, this is equivalent to gluing the two identified punctures of the sphere and obtaining a torus as a result:
\begin{equation}
    \begin{tikzpicture}[baseline={([yshift=-0.5ex]current bounding box.center)}]
        \draw [fill=black!10] (0,0) circle (1);
        \draw [fill=white] (0,0.8) ellipse (0.25 and 0.1);
        \draw [dashed] (0,-0.8) ellipse (0.25 and 0.1);
        
        \draw [dashed] (0,0) ellipse (1 and 0.4); 

        \draw (0,0.8) -- (0,0.352);
        \draw [dotted] (0,0.352) -- (0,-0.032);
        \draw (0,-0.032) -- (0,-0.8);
        \draw (-0.64, 0.48) node [circle,fill=black,inner sep=1pt] {} -- (0,0.48);
        \draw (-0.784, -0.16) node [circle,fill=black,inner sep=1pt] {} -- (0,-0.16);
        \draw (-0.64, -0.48) node [circle,fill=black,inner sep=1pt] {} -- (0,-0.48);
        \node at (0,0.8) [circle,fill=red,inner sep=1pt] {};
        \node at (0,-0.8) [circle,fill=red,inner sep=1pt] {};
    \end{tikzpicture}\quad  \cong \quad 
    \begin{tikzpicture}[baseline={([yshift=-0.5ex]current bounding box.center)}]
        \draw [fill=black!10] (-0.4,0.8) rectangle (0.4,-0.8);
        \draw [fill=white] (0,0.8) ellipse (0.4 and 0.1);
        \draw [dashed, fill=black!10] (0,-0.8) ellipse (0.4 and 0.1);

        \draw (0,0.8) -- (0,0.352);
        \draw [dotted] (0,0.352) -- (0,-0.032);
        \draw (0,-0.032) -- (0,-0.8);
        \draw (-0.3, 0.48) node [circle,fill=black,inner sep=1pt] {} -- (0,0.48);
        \draw (-0.3, -0.16) node [circle,fill=black,inner sep=1pt] {} -- (0,-0.16);
        \draw (-0.3, -0.48) node [circle,fill=black,inner sep=1pt] {} -- (0,-0.48);
        \node at (0,0.8) [circle,fill=red,inner sep=1pt] {};
        \node at (0,-0.8) [circle,fill=red,inner sep=1pt] {};
    \end{tikzpicture}\quad  \cong \quad \begin{tikzpicture}[baseline={([yshift=-0.5ex]current bounding box.center)}]
        \draw [fill=black!10] (0,0) circle (1);
        \draw [fill=white] (0,0) circle (0.4);

        \draw (0,0) ++(225:0.55) arc (225:495:0.55);
        \draw[dotted] (0,0) ++(135:0.55) arc (135:225:0.55);
        \draw (0.8, 0) node [circle,fill=black,inner sep=1pt] {} -- (0.55,0);
        \draw (0, 0.8) node [circle,fill=black,inner sep=1pt] {} -- (0,0.55);
        \draw (0, -0.8) node [circle,fill=black,inner sep=1pt] {} -- (0,-0.55);
        \draw[dashed, rotate=45] (0.7,0) ellipse (0.3 and 0.12);
    \end{tikzpicture}\,.
\end{equation}
This recovers the description of $L$ anyons on a torus. Notice that the fusion ring we constructed lies inside the torus. A precise treatment would include a discussion about the inside and outside basis \cite{Pfeifer_PBC}; we do not pursue this here, except to note that these bases are related by a change of basis given by the unitary modular $S$-matrix. The corresponding state space is spanned by diagrams of the form
\begin{equation}
    \begin{tikzpicture}[
    baseline={([yshift=-0.5ex]current bounding box.center)},
    x=\cell cm,
    y=\cell cm
]
    \begin{scope}[shift={(0,0)}]
    \draw (0,0) ellipse (4.2 and 1.75);
    \draw (-3.35,1.05) -- (-3.35,3.05) node[above] {\scriptsize $a_1$};
    \draw (-1.40,1.65) -- (-1.40,3.35) node[above] {\scriptsize $a_2$};
    \draw ( 1.40,1.65) -- ( 1.40,3.35) node[above] {\scriptsize $a_{L-1}$};
    \draw ( 3.35,1.05) -- ( 3.35,3.05) node[above] {\scriptsize $a_L$};
    \node at (-4,1.2) {\scriptsize $x$};
    \node at (-2.4,1.9) {\scriptsize $x_1$};
    \node at ( 0.00,2.1) {\scriptsize $\cdots$};
    \node at ( 2.4,1.9) {\scriptsize $x_{L-1}$};
    \node at ( 4,1.2) {\scriptsize $x$};
    \node at (0,0) {$\otimes$};
\end{scope}
    \end{tikzpicture}
    \simeq
\begin{tikzpicture}[baseline={([yshift=-0.5ex]current bounding box.center)},x=\cell cm,y=\cell cm]
        \draw (0,0) node [above left] {\scriptsize $a_1$} -- (1.6,-1.6);
        \draw (2,0) node [above right] {\scriptsize $a_2$} -- (1,-1);
        \node [above right] at (4,0) {\scriptsize $\cdots$};
        \draw [thick, loosely dotted] (1.6,-1.6) -- (2.4,-2.4);
        \draw (6,0) node [above right] {\scriptsize $a_{L-1}$} -- (3,-3);
        \draw (8,0) node [above right] {\scriptsize $a_L$} -- (4,-4);
        \draw (2.4,-2.4) -- (4,-4) -- node[below right] {\scriptsize $\tilde{x}_{L-1}$} (2.5,-5.5) -- node[below right] {\scriptsize $x$} (1.5,-6.5);
        \draw (2.5,-5.5) -- (-3,0) -- (-4,-1) -- (1.5,-6.5);
        \node[below left] at (1.9,-1.2) {\scriptsize $\tilde{x}_1$};
        \node[below left] at (3.8,-3.1) {\scriptsize $\tilde{x}_{L-2}$};
        \node[below left] at (-0.5,-4.2) {\scriptsize $\overline{x}$};
        \node at (-3,-1) {$\otimes$};
    \end{tikzpicture}\,,
\end{equation}
where the $\otimes$ denotes the non-contractible loop and the two diagrams are related by a series of $F$-symbols~\cite{Pfeifer_PBC}\@. From the right picture, it is clear that the fusion space is $\bigoplus_xV_{\overline{x}a_1\dots a_L}^x\otimes V_{x\overline{x}}^0$, reflecting the fact that cutting the torus along a non-contractible cycle produces a cylinder, whose two boundaries carry dual charges.
Thus, the total charge of the entire system will always be $J=0$. Nevertheless, there is an additional superselection rule according to the anyonic charge $x$ threading the non-contractible loop. This superselection rule comes from a \emph{global} symmetry of the Hilbert space, which is represented by the topological symmetry operators $Y_\ell$, $\ell\in\mathcal{S}$, that correspond to fusing a loop of charge $\ell$ into the periodic chain~\cite{golden_chain, Pfeifer_PBC, Buican2017}\@
\begin{align}
    &\begin{tikzpicture}[
    baseline={([yshift=-0.5ex]current bounding box.center)},
    x=\cell cm,
    y=\cell cm
]
        \begin{scope}[shift={(0,0)}]
    \draw (0,0) ellipse (4.2 and 1.75);
    \draw (0,0) ellipse (2.8 and 1.00);
    \node at (0,0) {$\otimes$};
    \node at (2.95,-0.45) {\scriptsize $\ell$};
    \draw (-3.35,1.05) -- (-3.35,3.05) node[above] {\scriptsize $a_1$};
    \draw (-1.40,1.65) -- (-1.40,3.35) node[above] {\scriptsize $a_2$};
    \draw ( 1.40,1.65) -- ( 1.40,3.35) node[above] {\scriptsize $a_{L-1}$};
    \draw ( 3.35,1.05) -- ( 3.35,3.05) node[above] {\scriptsize $a_L$};
    \node at (-4,1.2) {\scriptsize $x_0$};
    \node at (-2.4,1.9) {\scriptsize $x_1$};
    \node at ( 0.00,2.1) {\scriptsize $\cdots$};
    \node at ( 2.4,1.9) {\scriptsize $x_{L-1}$};
    \node at ( 4,1.2) {\scriptsize $x_0$};
\end{scope}
    \end{tikzpicture}\nonumber\\
    &=\sum_{x_0',\dots,x_{L-1}'}\prod_{i=0}^{L-1} \big(F_{x_{i+1}'}^{a_i x_i\ell}\big)_{x_{i+1}x_i'}\,
    \begin{tikzpicture}[
    baseline={([yshift=-0.5ex]current bounding box.center)},
    x=\cell cm,
    y=\cell cm
]
        \begin{scope}[shift={(0,0)}]
    \draw (0,0) ellipse (4.2 and 1.75);
    \node at (0,0) {$\otimes$};
    \draw (-3.35,1.05) -- (-3.35,3.05) node[above] {\scriptsize $a_1$};
    \draw (-1.40,1.65) -- (-1.40,3.35) node[above] {\scriptsize $a_2$};
    \draw ( 1.40,1.65) -- ( 1.40,3.35) node[above] {\scriptsize $a_{L-1}$};
    \draw ( 3.35,1.05) -- ( 3.35,3.05) node[above] {\scriptsize $a_L$};
    \node at (-4,1.2) {\scriptsize $x_0$};
    \node at (-2.4,1.9) {\scriptsize $x_1$};
    \node at ( 0.00,2.1) {\scriptsize $\cdots$};
    \node at ( 2.4,1.9) {\scriptsize $x_{L-1}$};
    \node at ( 4,1.2) {\scriptsize $x_0$};
\end{scope}
    \end{tikzpicture}\,.\label{eq:topological_symmetry}
\end{align}

The set of $Y_\ell$ forms an algebra $Y_aY_b=\sum_{c}N^c_{ab}Y_c$ obeying the fusion rules. These operators fall under a class of non-invertible symmetries. The Hilbert space splits into a direct sum over eigenspaces of the $Y_\ell$. The resulting topological symmetry superselection rule in $x$ is global, and an analogous constraint does not exist in any local subsystem. Therefore, it does not affect the entanglement entropy. The only consequence is that any pure state can only be supported in one fixed $x$-sector, restricting the initial state of entanglement entropy calculations.

To compute the fusion space dimensions, we proceed in the same way as in Section~\ref{sec:large-L} for anyons on a disk. For the dimension of the full Hilbert space, we take the trace of the corresponding product of fusion matrices~\eqref{eq:fusion_matrix},
\begin{equation}
    \dim \mathcal{H}=\tr\left(\prod_{i=1}^L N_{a_i}\right)\,.
\end{equation}
Using the Verlinde formula~\eqref{eq:Verlinde_formula} and setting all physical anyons to $a_i=\mathfrak{j}$, this yields in terms of the modular $S$-matrix,
\begin{equation}
    \dim \mathcal{H} = \sum_{j\in\mathcal{S}} \left(\frac{S_{\mathfrak{j} j}}{S_{0j}}\right)^L.
\end{equation}
The local subsystem states again look like states for anyons on a disk~\cite[Section~4.2]{Bonderson_entanglement}\@. However, due to the periodic boundary condition of the torus, a local subsystem will no longer be restricted to have $0$ as its first intermediate fusion label $x_1$. Therefore, the dimension of a local subsystem of length $L_C$ and total charge $\gamma$ becomes,
\begin{equation}
    \dim V^\gamma_{a_1\dots a_{L_C}}=\left(\prod_{i=1}^{L_C} N_{a_i}\right)_{x_1\gamma}\,,
\end{equation}
which follows from altering Eq.~\eqref{eq:fusion_multiplicity_matrix} accordingly. Again, by setting all $a_i=\mathfrak{j}$ and using the Verlinde formula, we obtain
\begin{equation}
    \dim V^\gamma_{a_1\dots a_{L_C}}=\sum_{j\in\mathcal{S}} S_{x_1 j} S_{\gamma j}^* \left(\frac{S_{\mathfrak{j}j}}{S_{0j}}\right)^{L_C}.
\end{equation}

\section{Variance}\label{app:variance}
In this appendix, we outline the calculation of the anyonic variance $(\Delta\tilde{S}_A)^2_J$ provided in Eq.~\eqref{eq:exact_variance} in the main text. The derivation exactly parallels that of Ref.~\cite{BD}, but involves additional terms of the logarithm of the quantum dimensions. 

We first recall the AEE of a state in Eq.~\eqref{eq:state_AEE}
\begin{equation}
    \tilde{S}_A=H(\{p_\alpha\})+\sum_\alpha p_\alpha(H(\{\lambda_{\alpha,i}\})+\log(d_\alpha))\,.
\end{equation}
For the ease of notation, we define $S_{\alpha A}=H(\{\lambda_{\alpha,i})\}$, the Shannon entropy of the eigenvalues of $R_\alpha$. Similarly, the anyonic version $\tilde{S}_{\alpha A}=S_{\alpha A}+\log(d_\alpha)$ includes the topological $\log(d_\alpha)$ term. For completeness, the von Neumann entropy can be written $S_A=\tilde{S}_A-\sum_\alpha p_\alpha\log(d_\alpha)$. Because $\log(d_\alpha)$ is not a random variable,
\begin{equation}
\label{eq:AEE_vN}
    \langle\tilde{S}_{\alpha A}\rangle_J=\langle S_{\alpha A}\rangle_J+\log(d_\alpha)\,, \qquad (\Delta \tilde{S}_{\alpha A})^2_J=(\Delta S_{\alpha A})^2_J\,.
\end{equation}

In Ref.~\cite{BD}, the variance of the entanglement entropy was computed to be 
\begin{equation}
    (\Delta S_A)^2_J = \frac{1}{D_J+1}\left(\sum_\alpha \varrho_\alpha(\varphi_\alpha^2+\chi_\alpha) -\langle S_A\rangle^2_J \right)\,,
\end{equation}
with
\begin{align}
    \varrho_\alpha&=\frac{m_\alpha n_\alpha}{D_J}\,, \\
    \phi_\alpha&=\Psi(D_J+1)-\Psi(\max(m_\alpha,n_\alpha)+1) \nonumber \\
    &\phantom{={}}-\min\left(\frac{m_\alpha-1}{2n_\alpha},\frac{n_\alpha-1}{2m_\alpha}\right)\,,  \\
    \chi_\alpha&=\begin{cases}
        (m_\alpha+n_\alpha)\Psi'(n_\alpha+1) & \\ 
        \,\,-(D_J+1)\Psi'(D_J+1) & m_\alpha\leq n_\alpha\,,\\
        \,\,-\frac{(m_\alpha-1)(m_\alpha+2n_\alpha-1)}{4n_\alpha^2} \\[0.3cm]
        (m_\alpha+n_\alpha)\Psi'(m_\alpha+1) & \\ 
        \,\,-(D_J+1)\Psi'(D_J+1) & m_\alpha\geq n_\alpha\,.\\
        \,\,-\frac{(n_\alpha-1)(n_\alpha+2m_\alpha-1)}{4m_\alpha^2}
    \end{cases}
\end{align}
Because of the relations~\eqref{eq:AEE_vN}, the expression with $(\Delta \tilde{S}_A)^2_J$ is simply obtained by the replacement $\phi_\alpha\mapsto \varphi_\alpha=\phi_\alpha+\log(d_\alpha)$ and $\langle S_A\rangle_J\mapsto \langle \tilde{S}_A\rangle_J$. This leads to the variance expression~\eqref{eq:exact_variance} in the main text.

\section{$q$-symmetry resolved entanglement entropy}
\label{app:vN_entropy}
Aside from anyonic systems, the $U_q(\mathfrak{sl}_2)$ quantum group plays important roles in quantum systems as a $q$-deformation to $\SU(2)$ systems. For example, such spin networks in loop quantum gravity can be deformed as to include a cosmological constant~\cite{Major1996, Han2011, Dupuis2013, Dupuis2014, Dupuis2014a, Hahn2024}. In these systems, even though there is no topological content (\textit{e.g.\@} isotopy-invariance), there is still no notion of a local Hilbert space. It is thus appropriate to use the standard von Neumann entropy as a measure of bipartite entanglement. In this appendix we comment briefly on the entanglement in these systems. As discussed in Section~\eqref{sec:symmetry-resolved}, the quantity we calculate here is the quantum group generalization of the non-abelian symmetry-resolved entanglement entropy, which we call the $q$-deformed symmetry-resolved entanglement entropy. The following calculations are specific to the fusion category $\SU(2)_k$.

We start directly from the formulas for $\rho_\alpha$ and $\varphi_\alpha$ reported in the main text, but with the difference that there is no topological $\log(d_\alpha)$ term. Concretely, the von Neumann entropy is given by
\begin{align}
    \langle S_A\rangle_J&=\sum_\alpha \varrho_\alpha \varphi_\alpha\,, \quad \text{where} \quad \varrho_\alpha=\frac{m_\alpha n_\alpha}{D_J}\,, \nonumber \\
    \varphi_\alpha&=\Psi(D_J+1)-\Psi(\max(m_\alpha,n_\alpha)+1) \nonumber \\
    &\phantom{={}}-\min\left(\frac{m_\alpha-1}{2n_\alpha},\frac{n_\alpha-1}{2m_\alpha}\right)\,,
\end{align}
with asymptotic Hilbert space dimensions
\begin{align}
\begin{split}
    D_J&=(1+(-1)^{2\mathfrak{j}L+2J})S_{00}S_{J0}d_\mathfrak{j}^L(1+o(1))\,, \\
    m_\alpha&=(1+(-1)^{2\mathfrak{j}L_A+2\alpha})S_{00}S_{\alpha0}d_\mathfrak{j}^{L_A}(1+o(1))\,, \\
        n_\alpha&=(1+(-1)^{2\mathfrak{j}L_B+2(\alpha+J)})S_{\alpha0}S_{J0}d_\mathfrak{j}^{L_B}(1+o(1))\,.
\end{split}
\end{align}
We would like to explicitly perform the sum over $\alpha$ and obtain a closed-form expression. 

The expansion of $\varphi_\alpha$ in the main text gave
\begin{equation}
    \varphi_\alpha=\begin{cases}
        L_A\log(d_\mathfrak{j})-\log(d_\alpha)-\frac{1}{2d_J}\delta_{f,\frac{1}{2}}+o(1)\,, & f\leq1/2 \\
        L_B\log(d_\mathfrak{j})+\log(d_J)-\log(d_\alpha)+o(1)\,, & f>1/2
    \end{cases}\,.
\end{equation}
Similarly, the discrete probability distribution $\varrho_\alpha$ involves a ratio of dimensions.
\begin{equation}
    \varrho_\alpha=2S_{\alpha0}^2(1+o(1))=2\frac{d_\alpha^2}{\mathcal{D}^2}(1+o(1))\,,
\end{equation}
where again $\mathcal{D}=\sqrt{\sum_{a\in\mathcal{S}}d_a^2}$.

In the following, we will show results only for $f\leq1/2$. The $f>1/2$ case is obtained simply by adding a $\log(d_J)$ term and changing the volume law factor from $L_A$ to $L_B$. 

The entanglement entropy up to constant order is
\begin{equation}
    \langle S_A\rangle_J=L_A\log(d_\mathfrak{j})-\frac{2}{\mathcal{D}^2}\sum_\alpha d_\alpha^2\log(d_\alpha)-\frac{1}{2d_J}\delta_{f,\frac{1}{2}}\,. 
\end{equation}
The label $\alpha$ takes values over admissible spins $\{\alpha\mid 2\alpha\equiv2\mathfrak{j}L_A \pmod{2}\}$: all half-integer spins if $\mathfrak{j}$ is half-integer and $L_A$ is odd; all integer spins otherwise.

The remaining sum can be explicitly performed, and it is instructive to do so in order to compare with the analogous result in symmetry-resolved $\SU(2)$. We will need to examine the case $k$ odd/even separately. As a first step, it is useful to obtain an expression for the sum, where the sum is over all spins $\alpha\in\mathcal{S}$.

Begin by defining $\theta=\frac{\pi}{k+2}$ and $n=2\alpha+1$, and write
\begin{equation}
    \sum_{\alpha\in\mathcal{S}} d_\alpha^2\log(d_\alpha)=\frac{1}{\sin^2(\theta)}\sum_{n=1}^{k+1}\sin^2(n\theta)\log(\frac{\sin(n\theta)}{\sin(\theta)})\,,
\end{equation}
where the quantum dimensions for $\SU(2)_k$ are $d_\alpha=\sin(n\theta)/\sin(\theta)$. Splitting the logarithm gives one term which is evaluated using $\sum_{n=1}^{k+1}\sin^2(n\theta)=(k+2)/2$. The other term, after using $\sin^2(n\theta)=1/2(1-\cos(2n\theta))$, contains two pieces: $\sum_{n=1}^{k+1}\log(\sin(n\theta))=\log(\prod_{n=1}^{k+1}\sin(n\theta))=\log(k+2)-(k+1)\log(2)$, and $\sum_{n=1}^{k+1}\cos(2n\theta)\log(\sin(n\theta))=\log(k+2)+\log(2)+\frac{\pi}{2}\cot(\theta)+\Psi(\frac{1}{k+2})+\gamma_E$, known as the Gauss digamma theorem. Collecting all these terms, we have
\begin{widetext}
\begin{equation}
    \sum_{\alpha\in\mathcal{S}}d_\alpha^2\log(d_\alpha)=-\frac{1}{2\sin^2(\theta)}\left[(k+2)\log(2\sin(\theta))+\frac{\pi}{2}\cot(\theta)+\Psi\left(\frac{1}{k+2}\right)+\gamma_E\right]
\end{equation}
We also note that $\mathcal{D}^2=1/S_{00}^2=(k+2)/(2\sin^2(\theta))$.

\subsection{$k$ odd}
When the level $k$ is odd, $j\mapsto k/2-j$ is an automorphism between integer and half-integer spins. Therefore 
\begin{equation}
    \sum_{\alpha\in S_{\text{int}}}d_\alpha^2\log(d_\alpha)=\sum_{\alpha\in S_{\text{h-int}}}d_\alpha^2\log(d_\alpha)=-\frac{1}{4\sin^2(\theta)}\left[(k+2)\log(2\sin(\theta))+\frac{\pi}{2}\cot(\theta)+\Psi\left(\frac{1}{k+2}\right)+\gamma_E\right]\,,
\end{equation}
and the average entanglement entropy has the expansion
\begin{equation}
\label{eq:SU(2)_k_SREE}
    \langle S_A\rangle_J=f L\log(d_\mathfrak{j})+\log(2\sin(\theta))+\frac{\pi}{2(k+2)}\cot(\theta)+\frac{1}{k+2}\Psi\left(\frac{1}{k+2}\right)+\frac{\gamma_E}{k+2}-\frac{1}{2d_J}\delta_{f,\frac{1}{2}}+o(1)\,.
\end{equation}
This equation describes the average $U_q(\mathfrak{sl}_2)$ symmetry-resolved entanglement entropy. When $J=0$, this expression is directly comparable to Eq.~(228)\footnote{The published version contains a typo in this equation, see Ref.~\cite{EESU(2)1.1}.} of Ref.~\cite{SREESU(2)} which describes the average $\SU(2)$ symmetry-resolved entanglement entropy:
\begin{equation}
\label{eq:SU(2)_SREE}
    \langle S_{GA}\rangle_0=fL\log(2)-\frac{1}{2}\log(L)-\frac{3(f+\log(1-f))}{2}-\frac{1}{2}\log\left(\frac{\rme^{2-\gamma_E}f(1-f)}{2}\right)-\frac{1}{2}\delta_{f,\frac{1}{2}}+o(1)\,.
\end{equation}
As discussed in the main text, the structure of UMTCs, and in particular fusion categories, fixes Hilbert space dimensions to growing exponentially without a polynomial prefactor. This leads to expression~\eqref{eq:SU(2)_k_SREE} having no $\log(L)$ and $O(1)$ $f$-dependent terms. The naïve $k\to\infty$ limit of Eq.~\eqref{eq:SU(2)_k_SREE} is ill-defined because, as discussed in Section~\ref{sec:large-L}, the approximations made in the Verlinde formula fail. The careful treatment results in Eq.~\eqref{eq:SU(2)_SREE}\@.

\subsection{$k$ even}
In the case where $k$ is even, we no longer expect the sum over the integer spins to equal the sum over the half-integer spins. The derivation of the result parallels that of the case above, so we just report the result below.  
\begin{align}
    \sum_{\alpha\in S_{\text{h-int}}}d_\alpha^2\log(d_\alpha)&=-\frac{1}{4\sin^2(\theta)}\left[(k+2)\log(2\sin(\theta))+\pi\cot(2\theta)+2\Psi\left(\frac{2}{k+2}\right)+2\gamma_E\right]\,, \\
    \sum_{\alpha\in S_{\text{int}}}d_\alpha^2\log(d_\alpha)&=-\frac{1}{4\sin^2(\theta)}\left[(k+2)\log(2\sin(\theta))+\pi(\cot(\theta)-\cot(2\theta))+2\left(\Psi\left(\frac{1}{k+2}\right)-\Psi\left(\frac{2}{k+2}\right)\right)\right]\,,
\end{align}
which gives
\begin{equation}
    \langle S_A\rangle_J=\begin{cases}
        fL\log(d_\mathfrak{j})+\log(2\sin(\theta))+\frac{\pi}{(k+2)}\cot(2\theta)+\frac{2}{k+2}\Psi\left(\frac{2}{k+2}\right)+\frac{2\gamma_E}{k+2}-\frac{1}{2d_J}\delta_{f,\frac{1}{2}}\,, \qquad \mathfrak{j}\in\mathbb{Z}+\frac{1}{2}, L_A\in2\mathbb{Z}+1\,, \\
        fL\log(d_\mathfrak{j})+\log(2\sin(\theta))+\frac{\pi}{k+2}(\cot(\theta)-\cot(2\theta))+\frac{2}{k+2}\left(\Psi\left(\frac{1}{k+2}\right)-\Psi\left(\frac{2}{k+2}\right)\right)-\frac{1}{2d_J}\delta_{f,\frac{1}{2}}\,, \qquad \text{otherwise.}
    \end{cases}
\end{equation}

\end{widetext}

\bibliography{references}

\end{document}